\documentclass{article} 
\usepackage{iclr2027_conference,times}
\usepackage{tabularx}
\usepackage[utf8]{inputenc} 
\usepackage[T1]{fontenc}
\usepackage{graphicx}
\usepackage{glossaries}
\usepackage{soul} 
\usepackage[bookmarksdepth=2]{hyperref} 
\usepackage{bbding}
\usepackage{float} 
\usepackage[many]{tcolorbox} 
\usepackage{enumitem} 
\usepackage{ifthen} 
\usepackage{amssymb} 
\usepackage{booktabs}
\usepackage{graphicx}
\usepackage{makecell}
\usepackage{amsmath}
\usepackage{url}
\usepackage[export]{adjustbox}
\usepackage{ragged2e} 
\usepackage{relsize} 
\usepackage{multirow} 
\usepackage{pbox} 
\usepackage{balance} 
\usepackage{moresize} 
\usepackage{xspace}
\usepackage{listings}
\usepackage{placeins} 
\usepackage{enumitem}

\newtcolorbox{mybox}[2][]{
top=0.15in,left=4pt,right=4pt,bottom=4pt,
fonttitle=\bfseries,
colbacktitle=gray,
colback=gray!5,
colframe=gray!40!black,
enhanced,
attach boxed title to top left={xshift=1.5em,yshift=-\tcboxedtitleheight/2},
boxed title style={size=small},
drop shadow={black!50!white},
title=#2,#1}

\newcommand{\countobservations}{
    \def \countobservations{1}
}
\newcounter{observation}
\countobservations

\newcommand{\countimplications}{
    \def \countimplications{1}
}
\newcounter{implication}
\countimplications

\newboolean{showcomments}
\setboolean{showcomments}{true}

\ifthenelse{\boolean{showcomments}}
{\newcommand{\nbc}[3]{
 {\colorbox{#3}{\bfseries\sffamily\scriptsize\textcolor{white}{#1}}}
 {\textcolor{#3}{\sf\small$\blacktriangleright$\textit{#2}$\blacktriangleleft$}}
 }
}
{\newcommand{\nbc}[3]{}
 }

\newcommand\ahsan[1]{\nbc{Ahsan}{#1}{blue}}
\newcommand\peter[1]{\nbc{Peter}{#1}{purple}}

\newcommand{\rabeya}[1]{\nbc{Rabeya}{#1}{forestgreen}}

\newcommand{\blackcircled}[1]{%
\tikz[baseline=(char.base)]{
  \node[shape=circle,fill=black,inner sep=1.1pt] (char)
  {\textcolor{white}{\footnotesize #1}};
}}

\usepackage{amsmath,amsfonts,bm}

\def\eqref#1{equation~\ref{#1}}

\def\1{\bm{1}}

\DeclareMathAlphabet{\mathsfit}{\encodingdefault}{\sfdefault}{m}{sl}
\SetMathAlphabet{\mathsfit}{bold}{\encodingdefault}{\sfdefault}{bx}{n}

\newcommand{\smalltt}[1]{\ifmmode{\mbox{\smaller\texttt{#1}}}\else{\smaller\tt #1}\fi}

\newcommand{\phead}[1]{\noindent {\bf #1}}
\newcolumntype{L}[1]{>{\raggedright\let\newline\\\arraybackslash\hspace{0pt}}m{#1}}
\newcolumntype{C}[1]{>{\centering\let\newline\\\arraybackslash\hspace{0pt}}m{#1}}
\newcolumntype{R}[1]{>{\raggedleft\let\newline\\\arraybackslash\hspace{0pt}}m{#1}}

\definecolor{promptbg}{RGB}{248,248,248}
\definecolor{promptframe}{RGB}{180,180,180}
\definecolor{forestgreen}{RGB}{34,139,34}
\lstdefinestyle{promptbox}{
    basicstyle=\ttfamily\scriptsize,
    breaklines=true,
    breakatwhitespace=false,
    frame=single,
    rulecolor=\color{promptframe},
    backgroundcolor=\color{promptbg},
    framesep=6pt,
    xleftmargin=4pt,
    xrightmargin=4pt,
    aboveskip=8pt,
    belowskip=8pt,
    showstringspaces=false,
    columns=fullflexible
}
\newcommand{\tool}{\textit{RETRACE}\xspace}

\title{\tool: From Entangled Repair Histories to Reusable Experience for CI Repair}

\author{
\textbf{Rabeya Khatun Muna}$^{1}$,
\textbf{Md Ahasanuzzaman}$^{1}$,
\textbf{Md Nakhla Rafi}$^{1}$,
\textbf{Yisen Xu}$^{1}$,
\textbf{Jinqiu Yang}$^{2}$,
\\
\textbf{Tse-Hsun (Peter) Chen}$^{1}$
\\[1ex]
$^{1}$Software Performance, Analysis, and Reliability (SPEAR) Lab,
Concordia University
\\
$^{2}$Concordia University
}
\iclrfinalcopy 
\begin{document}

\maketitle

\begin{abstract}
Large language model (LLM) agents increasingly reuse prior experience, but most approaches assume that problems and solutions are already aligned. Software histories rarely provide this alignment: a pull request (PR) may contain multiple continuous integration (CI) problems, failed attempts, reverted edits, and unrelated changes, obscuring which changes resolve each problem. We present \tool, a framework for reconstructing problem-level repair experience from such histories. \tool combines an endpoint view that reasons backward from changes retained in the passing revision with a development view that traces repair evolution forward through commit history. CI execution evidence reconciles the two views, and the recovered experience is represented at three abstraction levels, from concrete fixes to transferable repair patterns. For new failures, \tool retrieves relevant problem-level experience to guide repair. On \textsc{CI-Repair-Bench}, comprising 565 PR-level repairs from 101 repositories across 12 failure categories, \tool improves mini-SWE-agent Pass@1 from 19.6\% to 31.9\% with MiniMax-M2.5 and from 23.3\% to 32.8\% with DeepSeek-V4-Flash. On a matched subset, Codex improves from 15.5\% to 27.5\%. Combining both views consistently outperforms either alone, showing that recovering problem--change alignment enables historical CI repairs to serve as reusable repair experience.
\end{abstract}

\section{Introduction}
\label{sec:introduction}

Large language model (LLM) coding agents increasingly reuse prior solutions,
trajectories, and distilled strategies when solving new
tasks~\citep{ouyang2026reasoningbank,mu2026experepair,li2026conrad}. These approaches commonly begin with experience already organized around a known task: a known objective paired with a solution, or a trajectory with a clear objective and outcome. Software repositories record far more human problem solving, but not in this clean form. Their logs, commit histories, patches, and validation results arise as part of software development, without an explicit mapping between individual problems and the changes that resolved them.

In collaborative software development, proposed changes are commonly organized
as a pull request (PR), which groups one or more commits before they are
integrated into a repository. Continuous integration (CI) workflows assess
these changes by automatically compiling, testing, and analyzing the
repository~\citep{xu2025logsage}. When CI fails, developers may
revise the PR through several commits before the workflow passes. The resulting history records both the observed failures and the human changes that eventually resolved them, making completed PRs a potentially valuable source of repair experience.

This experience, however, is difficult to reuse. A PR may contain many commits that address different development tasks, including intermediate repair attempts, reverted changes, and work unrelated to the CI failure. The final patch records the net effect of all these changes, while the passing CI outcome shows only that the PR succeeded as a whole. The history is therefore organized at the PR level, but a repair agent needs guidance for an individual problem. Providing the entire PR adds substantial context without identifying which changes repaired that problem.

This difficulty is not unique to CI. Real-world histories often preserve the actions taken and the final outcome without recording how they are related. In CI repair, this relationship can be inferred from several artifacts. Failure logs and workflow definitions show what CI observed. The commit history shows how potential repairs were introduced and revised, while the endpoint patch shows which changes remained when CI passed. The broader challenge is how to learn from such noisy histories. In CI repair, this requires combining the available artifacts to link each problem to the changes that addressed it and the context in which the repair applies. 



We introduce \tool (\textbf{Re}pair experience through \textbf{C}hange \textbf{A}ttribution in \textbf{P}ull requests), a framework that reconstructs reusable, problem-level repair experience from entangled PR histories. 
For each historical PR, \tool first organizes the available CI execution evidence and then performs bidirectional change attribution through two views: 1) the endpoint view reasons backward from the changes retained in the passing revision to the CI problems they may address, and 2) the development view follows the commit sequence forward to trace how those changes were introduced, revised, or removed. \tool reconciles the two views to infer which developer changes addressed each CI problem and retains only associations supported by the historical artifacts. 
Each record describes the CI problem and its inferred cause, identifies the developer changes that addressed it and the files or symbols they modified, and retains the CI results and commits supporting the attribution.

A concrete repair record preserves implementation details, but these details may not transfer when repositories use different files, symbols, dependencies, or code structures. \tool therefore represents each repair at three levels of abstraction. The concrete experience preserves the developer changes linked to the problem; the contextual strategy describes how to recognize and repair similar failures; and the transferable pattern generalizes the failure mechanism and repair approach across repositories.
For a new CI repair, \tool queries all three levels separately for each observed problem. It then filters the retrieved records using the target CI evidence and current repository state before providing them to the repair agent. 



We study this problem using \textsc{CI-Repair-Bench}~\citep{muna2026ci}, which contains 565 PR-level CI repairs from 101 repositories and covers 12 failure categories. As shown in Table~\ref{tab:benchmark_statistics}, their patches modify, on average, $6.2\times$ as many files and $7.3\times$ as many lines as the reference patches in \textsc{SWE-bench}. Moreover, 473 instances (83.7\%) contain multiple failure types, increasing the likelihood that a PR-level history contains repair evidence relevant to more than one problem.


We evaluate \tool on \textsc{CI-Repair-Bench} using two repair agents and multiple LLM backbones against two baselines: repair without historical experience (\emph{No-Experience}) and direct reuse of historical repair plans (\emph{Historical-Plan}). \tool consistently achieves the highest Pass@1 across the evaluated settings. With mini-SWE-agent and MiniMax-M2.5, \tool reaches 31.9\% Pass@1,
outperforming \emph{No-Experience} and \emph{Historical-Plan} by +12.3 and +13.5 percentage points, respectively. Ablations further show that reconciling the endpoint patch and development history achieves the highest repair success, supporting the use of both views to reconstruct repair experience.


Our contributions are:
\begin{itemize}[leftmargin=*, noitemsep, topsep=2pt]
    \item We formulate learning from completed CI repair histories as a problem-repair attribution task. The task is to determine which developer changes addressed each CI problem using PR-level CI evidence, commit histories, and endpoint patches, while distinguishing those changes from unrelated development work.

    \item We introduce \tool, which reasons backward from the endpoint patch and forward through the commit history to reconstruct problem-level repair experience. \tool reconciles both views using CI execution evidence, represents each repair at three abstraction levels, and retrieves the resulting records separately for each target problem. 

    \item We assess \tool on \textsc{CI-Repair-Bench} using two repair agents and two LLM backbones. Compared with no historical experience, \tool improves Pass@1 from 19.6\% to 31.9\% with MiniMax-M2.5, from 23.3\% to 32.8\% with DeepSeek-V4-Flash, and from 15.5\% to 27.5\% with Codex on a matched subset. It also outperforms the historical-plan baseline, while the ablations show that combining the endpoint and development views achieves the highest Pass@1 across all settings.
\end{itemize}
\section{Related Work}

\begin{table*}[t]
\centering
\footnotesize
\setlength{\tabcolsep}{2.5pt}
\renewcommand{\arraystretch}{1.20}

\caption{Comparison of change size across repository-level software repair benchmarks.}
\label{tab:benchmark_statistics}

\begin{tabularx}{\textwidth}{
@{}l
*{8}{>{\centering\arraybackslash}X}
@{}
}
\toprule
\textbf{Metric (Avg.)}
& \shortstack{\textbf{CI-Repair-}\\\textbf{Bench}}
& \shortstack{\textbf{SWE-}\\\textbf{bench}}
& \shortstack{\textbf{SWE-bench}\\\textbf{Verified}}
& \shortstack{\textbf{SWE-}\\\textbf{Gym}}
& \shortstack{\textbf{SWE-}\\\textbf{bench Live}}
& \shortstack{\textbf{SWE-}\\\textbf{bench Pro}}
& \shortstack{\textbf{SWE-bench}\\\textbf{Multimodal}}
& \shortstack{\textbf{FEA-}\\\textbf{Bench}} \\
\midrule

\# changed files
& \textbf{10.5}
& 1.7
& 1.24
& 2.5
& 3.3
& 4.1
& 2.0\textsuperscript{\dag}
& 2.62 \\

\# changed lines
& \textbf{240.1}
& 32.8
& 14.32
& 69.8
& 102.6
& 107.4
& 27\textsuperscript{\dag}
& 128.5 \\

\bottomrule
\end{tabularx}

\vspace{2pt}
\begin{minipage}{\textwidth}
\scriptsize
\textit{Notes.}
The benchmarks are CI-Repair-Bench~\citep{muna2026ci},
SWE-bench~\citep{jimenez2024swe},
SWE-bench Verified~\citep{chowdhury2024swebenchverified},
SWE-Gym~\citep{swe_bench_gym},
SWE-bench-Live~\citep{zhang2026swe},
SWE-bench Pro~\citep{deng2025swe},
SWE-bench Multimodal~\citep{yang2024swe}, and
FEA-Bench~\citep{li-etal-2025-fea}.
\textsuperscript{\dag}SWE-bench Multimodal reports medians rather than
averages.
\end{minipage}
\end{table*}

\phead{Learning from historical repairs.} Prior LLM-based repair methods reuse bug--fix examples through retrieval~\citep{inferFix,wang2023rap}, reasoning demonstrations~\citep{yin2024thinkrepair}, stored experiences~\citep{mu2026experepair}, or reconstructed repair plans~\citep{yang2025lingxi,xu2026stair,li2026conrad}. Although their representations differ, these methods generally assume that historical repairs have already been separated into task-aligned examples, such as an issue paired with a fix, trajectory, or plan. CI histories do not naturally provide this alignment. A failing-to-passing pull request may span multiple commits, address several CI problems, and include intermediate or unrelated changes. \tool reconstructs this alignment by combining the endpoint patch, and the commit history to associate individual problems with their supporting repairs. To our knowledge, it is the first approach to reconstruct and reuse problem-level repair experience from multi-commit CI histories. 

\phead{Repository-level software engineering agents.}
Repository-level agents combine LLM reasoning with repository exploration, fault localization, code editing, execution, and iterative feedback~\citep{li2026advances,yang2024sweagent,wang2025openhands,zhang2024autocoderover,xia2024agentless,antoniades2025swesearch}. These systems primarily target SWE-bench-style issue resolution, where a GitHub issue specifies the problem and tests validate the patch. CI repair differs, where one failing workflow may contain multiple CI problems, and early failures hide downstream validation evidence. To evaluate this setting, we use CI-Repair Bench~\citep{muna2026ci}, a benchmark of CI repair instances whose changes are substantially larger than those in existing repository-level benchmarks (cf. Table~\ref{tab:benchmark_statistics}). 
\tool addresses this challenge by reconstructing problem-level repair records from completed CI repair histories. For a new CI failure, it retrieves relevant records and uses them with the observed CI logs, workflow configuration, and
current repository state to guide the repair.

\phead{CI failure understanding and diagnosis.}
Prior work uses CI execution evidence to identify informative log regions~\citep{brandt2020logchunks}, localize faults across failure types~\citep{chen2022useful,hassan2023uniloc}, classify failures~\citep{aidasso2026flaxifyer}, and generate diagnoses and repair suggestions~\citep{xu2025logsage}. These methods focus on failures visible in the current CI execution and do not reconstruct complete multi-problem repair tasks, in which earlier failures may prevent downstream validations from running and leave additional problems unobserved. \tool instead reconstructs problem-level repair experience from completed CI histories and uses it to guide a repository-level agent in resolving such
problems and making the entire CI workflow succeed.

\section{\tool: Learning from Entangled CI Repair Histories}
\label{sec:approach}

Reusing a completed CI repair requires determining which developer changes
addressed each CI problem. \tool treats this as a problem--repair attribution
task. The available artifacts play different roles in this attribution. The CI
logs and workflow definition describe the observed problems and their validation
context. The endpoint patch identifies the changes retained in the passing
revision, but may also contain unrelated work. The commit history shows how
changes were introduced, revised, or removed, but includes intermediate attempts
that did not remain in the final repair. \tool analyzes the endpoint
patch and commit history as complementary views and reconciles them using the CI
execution evidence. This produces problem-level repair records that can be reused to guide future repairs. 

\begin{figure}[t]
    \centering
    \includegraphics[width=0.95\textwidth]{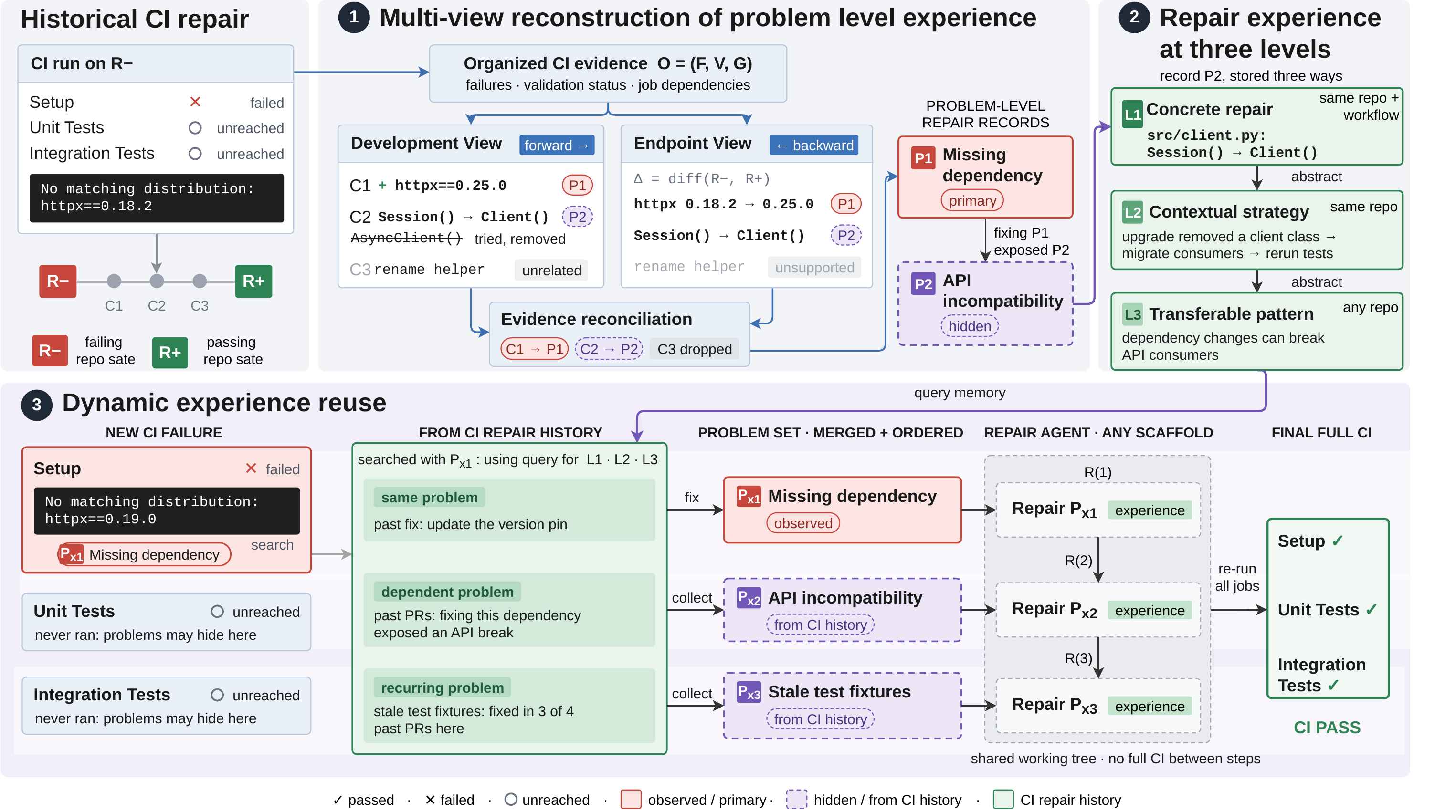}
    \caption{Overview of \tool.}
    \label{fig:overall-approach}
\end{figure}

Figure~\ref{fig:overall-approach} summarizes the framework. \tool first organizes
the CI execution evidence and performs bidirectional change attribution through
the endpoint and development views
(\S\ref{sec:experience-induction}). It then represents each reconstructed record
at three abstraction levels: concrete repair experience, a contextual repair
strategy, and a transferable repair pattern
(\S\ref{sec:experience-representation}). For a target repair, \tool retrieves
from the three levels separately for each target problem, filters the retrieved
records using the target CI evidence and current repository state, and provides
the retained experience to the repair agent
(\S\ref{sec:experience-reuse}).

We represent a completed historical repair as
\begin{equation}
    h_i =
    \left(
        R_i^{-},
        E_i,
        C_i,
        \Delta_i,
        R_i^{+}
    \right),
    \label{eq:historical_repair}
\end{equation}

where $R_i^{-}$ and $R_i^{+}$ denote the failing and CI-successful repository states, respectively. $E_i$ denotes the available CI evidence, including the associated execution logs and workflow information. $C_i=\langle c_{i1},\ldots,c_{im_i}\rangle$ is the sequence of developer commits between the two states and $\Delta_i=\operatorname{diff}(R_i^{-},R_i^{+})$ is the cumulative patch.


These artifacts play different roles in reconstruction. The CI evidence $E_i$ describes the observed failures and their workflow context and anchors both views. The \emph{endpoint view}, derived from $\Delta_i$, reasons backward from the changes retained in the passing revision to the CI problems they may have addressed. $\Delta_i$ contains the net changes that may include work unrelated to the CI problems. The \emph{development view}, derived from $C_i$, follows the repair through the commit sequence and shows when candidate changes were introduced, modified, or removed. It captures the repair's development but also includes incomplete or abandoned attempts. The endpoint view narrows the candidate repairs to changes that remained, while the development view provides their history. \tool reconciles the candidates from both views against $E_i$ to infer the problem--repair associations.

Given a collection of completed CI repair histories, $\mathcal{H}=\{h_1,\ldots,h_n\}$, \tool reconstructs
\begin{equation}
    \mathcal{Z}
    =
    \operatorname{Reconstruct}(\mathcal{H})
    =
    \bigcup_{i=1}^{n}\mathcal{Z}_i,
    \qquad
    \mathcal{Z}_i
    =
    \left\{
        z_{i1},\ldots,z_{ik_i}
    \right\},
    \label{eq:experience_induction}
\end{equation}
where $k_i$ is the number of problem-level repair records reconstructed from
$h_i$. Each $z_{ij}$ links one CI problem to the developer changes inferred to
have addressed it, together with the affected artifacts, relevant validation
evidence, and historical provenance. Reconstructing these records requires
\tool to identify relevant repair evidence, separate it from unrelated
development work, and infer which changes contributed to resolving each
problem. Here, \emph{learning} refers to constructing, abstracting, and reusing
these records rather than updating the parameters of the underlying LLM.


\subsection{Multi-View Reconstruction of Problem-Level Experience}
\label{sec:experience-induction}
The central difficulty is to infer which developer changes addressed each CI
problem. For each historical repair, \tool first organizes the raw CI evidence. It then uses the resulting organized CI evidence to guide
bidirectional change attribution through two complementary views. The endpoint
view reasons backward from the changes retained in the passing revision to the problems they may have addressed. The development view traces changes forward through the commit history to determine how potential repairs were introduced,
revised, or removed. Both views produce candidate problem--repair associations, which \tool reconciles using the CI evidence to construct the final problem-level repair records.

\phead{Organizing the CI evidence.}
Raw CI logs contain setup output, repeated messages, interleaved commands, and errors propagated from earlier steps. To make this evidence usable for attribution, \tool aligns the log output with the jobs, steps, and commands defined in the CI workflow. It extracts execution status, diagnostic messages, referenced artifacts, and environment information. It also examines the workflow definition and referenced scripts to identify dependencies, execution conditions, and validation commands. We represent the organized CI evidence as
\begin{equation}
    O_i
    =
    \operatorname{Organize}(E_i)
    =
    \left(
        \mathcal{F}_i,
        \mathcal{V}_i,
        \mathcal{G}_i
    \right),
    \label{eq:ci_observation}
\end{equation}

where $\mathcal{F}_i$ contains the observed failures, $\mathcal{V}_i$ records the status and context of each validation, and $\mathcal{G}_i$ captures dependencies among jobs and steps. Each failure retains its diagnostic message, originating command, workflow location, and referenced artifacts. Validations are marked as successful, failed, or unreached.

\phead{Recovering repairs from the endpoint view.}
The cumulative patch $\Delta_i$ shows the changes between the failing and
passing revisions, but not which changes resolved each problem. \tool splits
the patch into file-level and hunk-level changes and groups related changes using
code and workflow relations, such as imports, caller-callee links,
test-source links, and configuration-consumer links.

For each group, an LLM examines the changed code together with the failures, validations, and workflow information in $O_i$. If the CI evidence supports a connection between the changes and an observed problem, \tool creates a candidate problem-repair record. Changes with no supported connection to a CI problem are excluded. Each candidate has the form
\begin{equation}
    z_{ij}
    =
    \left(
        p_{ij},
        a_{ij},
        r_{ij},
        v_{ij},
        e_{ij}
    \right),
    \label{eq:experience_record}
\end{equation}
where $p_{ij}$ describes the problem and its inferred cause, $a_{ij}$ records the developer repair, $r_{ij}$ captures the affected artifacts and relevant repository context, $v_{ij}$ contains the validation evidence, and $e_{ij}$ records the historical evidence supporting the association between the problem and repair. Each record captures one failure mechanism and its repair, even when the repair spans multiple files or commits. \tool retains a record only when the historical artifacts support both the inferred cause and the attributed repair.

\phead{Tracing repairs through the development view.}
The endpoint patch contains only changes retained in $R_i^{+}$. To capture how the repair developed, \tool analyzes
$C_i=\langle c_{i1},\ldots,c_{im_i}\rangle$ in chronological order. For each commit, an LLM examines its changes together with $O_i$ and identifies which observed CI problems they address. Changes with no supported connection to a CI problem are excluded. Changes that address the same problem across multiple commits are combined into one record, while changes for different problems remain separate. Thus, one record may span several commits, and one commit may contribute to several records. Changes that are later revised or removed are kept as development history but are not included in the final repair unless their effect remains in
$R_i^{+}$. If intermediate commits are unavailable, \tool relies only on the endpoint view.

\phead{Reconciling evidence across views.}
The endpoint and development analyses may produce records for the same problem.
\tool first identifies records to compare based on their problem description,
affected artifacts, failure type, and validation. An LLM then classifies each pair as \emph{equivalent},
\emph{complementary}, \emph{conflicting}, or \emph{distinct}. Equivalent
records are consolidated. Complementary records are merged when they describe
different parts of the same problem and repair. Distinct records remain
separate. Conflicting records are checked against $O_i$, $C_i$, $\Delta_i$,
and $R_i^{+}$.

For each retained field, \tool requires the model to identify supporting historical evidence. During reconciliation, fields for which no supporting artifact can be identified are removed. Records found in only one view may still be retained when the available CI and repository evidence supports their problem-repair association. The resulting set $\mathcal{Z}_i$ contains the problem-repair records reconstructed from historical repair $h_i$. 

\subsection{Repair Experience at Three Levels of Abstraction}
\label{sec:experience-representation}

Concrete repair records are useful when a new failure occurs in a similar
repository context, but they may not transfer well when files, symbols,
dependencies, or implementations differ. More abstract representations can
apply across such differences, but contain fewer implementation details.
\tool therefore stores three linked representations of each repair record:
\begin{equation}
    \mathbf{z}_{ij}
    =
    \left(
        z_{ij}^{\mathrm{L1}},
        z_{ij}^{\mathrm{L2}},
        z_{ij}^{\mathrm{L3}}
    \right),
    \label{eq:three_level_experience}
\end{equation}
where $z_{ij}^{\mathrm{L1}}$ is the concrete repair experience,
$z_{ij}^{\mathrm{L2}}$ is a contextual repair strategy, and
$z_{ij}^{\mathrm{L3}}$ is a transferable repair pattern. All three
representations refer to the same historical repair and remain linked to its
supporting evidence.

\phead{L1: Concrete repair experience.}
The concrete record preserves the observed failure, inferred cause, affected
files and symbols, developer changes, validation, and commit evidence from
$z_{ij}$. It includes only the changes attributed to that problem, rather than
the entire pull request. This preserves the implementation details of the
historical repair while excluding unrelated changes.

\phead{L2: Contextual repair strategy.}
An LLM abstracts the concrete record into a repair strategy for similar repository contexts by replacing specific file paths, symbol names, and literal edits with the roles they play and the repair operations performed on them. The resulting strategy describes the failure mechanism, diagnostic signals, relationships among affected artifacts, repair steps, validation procedure, and any known pitfalls. Details such as dependency versions, configuration requirements, API assumptions, and workflow structure are kept only when they determine whether the strategy applies.

\phead{L3: Transferable repair pattern.}
An LLM further abstracts the repair into a pattern that can apply across
repositories. The pattern removes repository-specific identifiers and
implementation details while preserving the general failure mechanism,
diagnostic signals, artifact roles, repair principles, applicability
conditions, and historically observed follow-on problems.

The higher-level representations must remain supported by the lower-level
repair evidence and do not introduce new repair information. They are derived
as
\begin{equation}
    z_{ij}^{\mathrm{L2}}
    =
    \operatorname{Abstract}_{2}
    \left(
        z_{ij}^{\mathrm{L1}}
    \right),
    \qquad
    z_{ij}^{\mathrm{L3}}
    =
    \operatorname{Abstract}_{3}
    \left(
        z_{ij}^{\mathrm{L1}},
        z_{ij}^{\mathrm{L2}}
    \right).
    \label{eq:three_level_abstraction}
\end{equation}

\tool organizes the reconstructed problem-level repair records by abstraction level:

\begin{equation}
    \mathcal{M}^{(\ell)}
    =
    \left\{
        z_{ij}^{(\ell)}
        \mid
        h_i \in \mathcal{H},
        1 \leq j \leq k_i
    \right\},
    \qquad
    \ell \in
    \left\{
        \mathrm{L1},
        \mathrm{L2},
        \mathrm{L3}
    \right\}.
    \label{eq:three_level_memory}
\end{equation}
This allows \tool to retrieve concrete historical repairs, contextual repair
strategies, and more general repair patterns depending on the target problem.

\subsection{Dynamic Experience Reuse for Target Repair}
\label{sec:experience-reuse}

A target CI failure may involve multiple problems, including problems hidden because earlier failures prevented downstream validations from running. Retrieving a single repair experience for the entire failure can conflate these problems and overlook relevant historical evidence. \tool therefore constructs a target-conditioned problem set using the observed failure and historical dependencies, and a recurring-problem set from prior repairs in the same repository and workflow. It merges and orders both sets, then provides problem-specific historical guidance to the repair agent.

\phead{Target-conditioned problem set.}
For a target repair $x$, \tool first organizes the failing CI evidence as $O_x=\operatorname{Organize}(E_x)$. An LLM decomposes $O_x$ into a set of atomic problems $\mathcal{P}_x^{\mathrm{atomic}}$, where each problem records its
failure signals, inferred cause, affected artifacts, validation command, and relevant repository and workflow context.

For each atomic problem, \tool constructs a separate query for each
abstraction level and independently retrieves the top-$K$ most similar
candidate historical problem--repair records:

\begin{equation}
    \mathcal{C}_{xj}^{(\ell)}
    =
    \operatorname{TopK}_{z \in \mathcal{M}^{(\ell)}}
    \operatorname{sim}
    \left(
        q_{xj}^{(\ell)},
        z
    \right),
    \qquad
    \ell \in
    \left\{
        \mathrm{L1},
        \mathrm{L2},
        \mathrm{L3}
    \right\},
    \label{eq:level_specific_retrieval}
\end{equation}
where $q_{xj}^{(\ell)}$ is the query for retrieving historical repair experience at abstraction level $\ell$ and
$\operatorname{sim}(\cdot,\cdot)$ is cosine similarity. 
Because semantic similarity does not guarantee applicability, an
LLM further compares each candidate with the target evidence and retains only experience whose failure mechanism, repair conditions, and assumptions match the target. 

Since resolving one problem may expose a downstream problem, \tool uses dependency relations from selected historical experience to expand the problem set. During historical problem-repair experience reconstruction, \tool retains a relation $p_a\rightarrow p_b$ when the historical evidence indicates that resolving $p_a$ is a prerequisite for $p_b$. When a target problem matches $p_a$, its dependent problem $p_b$ and associated repair experience are added as a candidate problem. The initially decomposed problems ($\mathcal{P}_x^{\mathrm{atomic}}$) and dependency-derived problems together form the target-conditioned set $\mathcal{P}_x^{\mathrm{target}}$.

\phead{Recurring-problem set.}
To account for common repository- and workflow-specific problems that may not be connected to the observed failure by an explicit dependency, \tool also constructs a recurring-problem set from historical repairs. A problem is considered recurring when similar reconstructed problems appear across multiple repairs in the same repository and workflow, based on their descriptions, root causes, affected artifacts, failure types, and repair strategies. For a new
target in that repository and workflow, each recurring problem is added to $\mathcal{P}_x^{\mathrm{rec}}$ as a candidate problem, and its supporting historical repair experience is retained as guidance.


\phead{Consolidation and ordering.}
Before repair begins, \tool combines the two problem sets:
\begin{equation}
    \mathcal{P}_x
    =
    \operatorname{Merge}
    \left(
        \mathcal{P}_x^{\mathrm{target}}
        \cup
        \mathcal{P}_x^{\mathrm{rec}}
    \right).
    \label{eq:final_problem_set}
\end{equation}

Here, $\operatorname{Merge}$ combines the two sets and consolidates equivalent problems while preserving their evidence and associated repair experience. \tool then orders the problems using target-workflow dependencies and historical prerequisite relations. Unconstrained problems follow their CI execution order, followed by their associated hypotheses. This produces the final sequence of problems $P_x=\langle p_{x1},\ldots,p_{xN_x}\rangle$ for the target repair $x$.

\phead{Sequential repair and validation.}
\tool processes $P_x$ sequentially in a shared working tree. At step $j\in{1,\ldots,N_x}$, the repair agent updates the current repository state using the $j$-th problem its CI evidence, and its associated historical repair guidance:

\begin{equation}
R_x^{(j+1)}
=
\operatorname{Repair}
\left(
R_x^{(j)},p_{xj},O_{xj},G_{xj}
\right),
\label{eq}
\end{equation}
where $R_x^{(j)}$ is the repository state before repairing $p_{xj}$,
$O_{xj}$ is its problem-specific CI evidence, and $G_{xj}$ is its associated historical repair guidance. The initial state $R_x^{(1)}$ is the failing target revision, and each subsequent state retains the changes made for all preceding problems. The agent may run problem-specific validation commands when feasible, but \tool never re-executes the complete CI workflow between repair steps. After all $N_x$ problems have been processed, the final state $R_x^{(N_x+1)}$ is evaluated by rerunning the complete CI workflow. The repair is successful only if every CI job and validation passes. Additional implementation details are provided in
Appendix~\ref{app:dynamic-reuse}.

\section{Experiments}
\label{sec:experiment}
\subsection{Experimental Setup}
\label{sec:experimental-settings}

\phead{Benchmark and historical repairs.}
We evaluate \tool on \textsc{CI-Repair-Bench}~\citep{muna2026ci}, comprising
565 PR-level CI repair instances from 101 repositories and 12 failure
categories. Unlike \textsc{SWE-CI}~\citep{chen2026sweci}, which evaluates
long-term maintainability through incremental feature development over 100
tasks, \textsc{CI-Repair-Bench} targets CI failure repair: 83.7\% of instances
contain multiple concurrent failure types requiring heterogeneous repairs, and
related repair evidence recurs within and across repositories. For repositories
with multiple instances, the earliest 30\% form the historical set, yielding
157 historical repairs and 408 evaluation targets. Appendix~\ref{app:benchmark}
provides further benchmark and split details.

\phead{Models and repair agents.}
We assess three configurations: mini-SWE-agent~\citep{minisweagent} with MiniMax-M2.5, mini-SWE-agent with DeepSeek-V4-Flash, and
Codex~\citep{codex} with MiniMax-M2.5. Both mini-SWE-agent configurations are run on all 408 targets. Because of its higher inference cost, Codex is run on a randomly selected subset of 200 targets. To compare the two repair agents, we report the mini-SWE-agent results on the same 200 targets, keeping the model and target instances fixed. Appendix~\ref{app:experimental-details} provides the detailed model and agent settings.

\phead{Baselines.} The \emph{No-Experience} baseline uses the same configuration settings as  \tool but receives no historical experience.  Directly providing complete PR histories is impractical because their length can exceed the agent's context window. Hence, we introduce a plan-based baseline, \emph{Historical-Plan}, which provides the mini-SWE-agent with compact repair plans generated from the same historical repairs following ThinkRepair~\citep{yin2024thinkrepair}. We choose ThinkRepair because, like \tool, it learns from historical bug–fix examples rather than the agent’s own trajectories (e.g., ExpeRepair~\citep{mu2026experepair}), so both methods use the same historical repairs. Note that both \emph{Historical-Plan} and \tool use the historical example instances.

\phead{Evaluation.}
We use Pass@1 as the primary end-to-end repair metric, measuring the percentage
of instances repaired on the first attempt. Each generated patch is applied to
the failing revision and evaluated by rerunning the complete GitHub Actions
workflow; an instance is resolved only if the workflow passes.
Appendix~\ref{app:experimental-details} provides detailed settings and prompts.

\subsection{Results}


\phead{Performance of \tool.} \tool improves CI repair across both LLM backbones in mini-SWE-agent, raising \textit{\textbf{Pass@1 from 19.6\% to 31.9\% with MiniMax-M2.5}} and \textit{\textbf{from 23.3\% to 32.8\% with DeepSeek-V4-Flash.}} As shown in Table~\ref{tab:main-results}, these correspond to gains of 12.3 and 9.5 percentage points over \emph{No-Experience}, respectively; both differences remain significant after Holm correction
($p_{\mathrm{Holm}}<0.05$). Using the same historical repair instances, \textit{\textbf{\tool also outperforms \emph{Historical-Plan} by  13.5 and 14.9 percentage points}}. In contrast, \emph{Historical-Plan} performs below \emph{No-Experience} under both backbones, supporting that problem--repair reconstruction is essential before reuse and using raw experiences adds too much noise to the context. The improvement also extends across repair agents: on the 200-instance subset, \textit{\textbf{\tool raises mini-SWE-agent from 23.0\% to 27.5\% and Codex from 15.5\% to 27.5\%.}} These results show that the benefit of problem-level repair experience holds across both evaluated LLM backbones and repair-agent scaffolds.


\begin{table*}[t]
\centering
\caption{Main results on \textsc{CI-Repair-Bench}.
$^{*}$ indicates that \tool significantly outperforms the marked baseline
(exact McNemar test with Holm correction; $p_{\mathrm{Holm}}<0.05$).}
\label{tab:main-results}

\setlength{\tabcolsep}{3pt}
\renewcommand{\arraystretch}{0.95}

\begin{tabular}{@{}clllccc@{}}
\toprule
\textbf{Eval.} &
\textbf{Agent} &
\textbf{Model} &
\textbf{Method} &
\textbf{Pass@1 (\%)} &
\textbf{Input (M)} &
\textbf{Cost (\$)} \\
\midrule

\multirow{6}{*}{%
\rotatebox[origin=c]{90}{\shortstack{Full\\($n=408$)}}}
& \multirow{3}{*}{mini-SWE}
& \multirow{3}{*}{MiniMax-M2.5}
& No-Experience
& 19.6$^{*}$
& 1.081
& 0.080 \\

&&& Historical-plan
& 18.4$^{*}$
& 1.144
& 0.050 \\

&&& \tool
& \textbf{31.9}
& 1.903
& 0.139 \\

\cmidrule(lr){2-7}

& \multirow{3}{*}{mini-SWE}
& \multirow{3}{*}{DeepSeek-V4-Flash}
& No-Experience
& 23.3$^{*}$
& 1.023
& 0.033 \\

&&& Historical-plan
& 17.9$^{*}$
& 1.282
& 0.022 \\

&&& \tool
& \textbf{32.8}
& 1.800
& 0.052 \\

\midrule

\multirow{4}{*}{%
\rotatebox[origin=c]{90}{\shortstack{Subset\\($n=200$)}}}
& \multirow{2}{*}{mini-SWE}
& \multirow{2}{*}{MiniMax-M2.5}
& No-Experience
& 23.0
& 1.081
& 0.080 \\

&&& \tool
& \textbf{27.5}
& 1.463
& 0.113 \\

\cmidrule(lr){2-7}

& \multirow{2}{*}{Codex}
& \multirow{2}{*}{MiniMax-M2.5}
& No-Experience
& 15.5$^{*}$
& 3.128
& 0.142 \\

&&& \tool
& \textbf{27.5}
& 2.210
& 0.118 \\

\bottomrule
\end{tabular}

\vspace{1mm}
\parbox{\textwidth}{\footnotesize\raggedright
\textit{Note:} Input denotes the average input-token usage in millions per
instance; Cost denotes the average inference cost per instance.}
\end{table*}
\phead{Cost Analysis.} With mini-SWE-agent, \tool adds a small absolute increase in inference cost
but produces substantially better repair results. With MiniMax-M2.5, the
average cost increases by \$0.059 per instance, from \$0.080 to \$0.139, \textit{\textbf{while
Pass@1 improves by 12.3 percentage points}}. With DeepSeek-V4-Flash, the cost
increases by \$0.019, from \$0.033 to \$0.052, alongside a \textit{\textbf{9.5-point
improvement}}. With Codex, \tool improves Pass@1 while also reducing cost from
\$0.142 to \$0.118 per instance. Its average input usage decreases from 3.13M
to 2.21M tokens, consistent with the retrieved experience helping limit
unnecessary exploration. Note that these are end-to-end costs that include memory retrieval, reuse, and the agent's planning and execution. 

\phead{Ablation Studies.} Both reconstruction views and higher abstraction levels contribute to \tool's repair performance. \textit{\textbf{The full reconstruction reaches 31.9\%, 32.8\%, and 27.5\% Pass@1, outperforming both single-view variants}}, as shown in Table~\ref{tab:reconstruction}. Removing either the endpoint or development view lowers Pass@1, with the full method exceeding the stronger single-view variant by 3.5, 4.4, and 2.0 percentage points, respectively. Table~\ref{tab:memory} shows a similar progression across abstraction levels: \textit{\textbf{Pass@1 increases from 26.0\% with L1 to 28.0\% with L1+L2 and 30.5\% with L1+L2+L3}}, for a total gain of 4.5 points over L1. We conduct the abstraction-level ablation with mini-SWE-agent and DeepSeek-V4-Flash rather than Codex because Codex is more expensive to evaluate and provides a less favorable cost-performance setting for additional ablation runs.

\begin{table*}[t]
\centering
\caption{Effect of reconstruction strategy on \textsc{CI-Repair-Bench}.
All configurations use three-level experience abstractions.
$^{*}$ indicates that \tool$_{\mathrm{Full}}$ significantly outperforms the marked single-view variant (exact McNemar test with Holm correction; $p_{\mathrm{Holm}}<0.05$).}
\label{tab:reconstruction}

\setlength{\tabcolsep}{2.5pt}
\renewcommand{\arraystretch}{0.95}

\begin{tabular}{@{}lllccc@{}}
\toprule
\textbf{Agent ($n$)} &
\textbf{Model} &
\textbf{Configuration} &
\textbf{Pass@1 (\%)} &
\textbf{Input (M)} &
\textbf{Cost (\$)} \\
\midrule

\multirow{3}{*}{mini-SWE (408)}
& \multirow{3}{*}{MiniMax-M2.5}
& \tool$_{\textit{w/oEndpoint}}$
& 28.4*
& 1.262
& 0.108 \\

&& \tool$_{\textit{w/oDevelopment}}$
& 27.2*
& 1.565
& 0.127 \\

&& \tool$_{\textit{Full}}$
& \textbf{31.9}
& 1.903
& 0.139 \\

\midrule

\multirow{3}{*}{mini-SWE (408)}
& \multirow{3}{*}{DeepSeek-V4-Flash}
& \tool$_{\textit{w/oEndpoint}}$
& 28.4*
& 1.194
& 0.040 \\

&& \tool$_{\textit{w/oDevelopment}}$
& 27.7*
& 1.480
& 0.048 \\

&& \tool$_{\textit{Full}}$
& \textbf{32.8}
& 1.800
& 0.052 \\

\midrule

\multirow{3}{*}{Codex (200)}
& \multirow{3}{*}{MiniMax-M2.5}
& \tool$_{\textit{w/oEndpoint}}$
& 25.5
& 1.671
& 0.095 \\

&& \tool$_{\textit{w/oDevelopment}}$
& 23.5*
& 2.632
& 0.136 \\

&& \tool$_{\textit{Full}}$
& \textbf{27.5}
& 2.210
& 0.118 \\

\bottomrule
\end{tabular}

\vspace{1mm}
\parbox{\textwidth}{\footnotesize\raggedright
\textit{Note:} Input and Cost denote the average per-instance input-token
usage in millions and inference cost, respectively.}
\end{table*}

\begin{table}[t]
\centering
\caption{Ablation of repair-experience abstraction levels on a fixed
200-instance subset using mini-SWE-agent with DeepSeek-V4-Flash.
All other \tool components are held fixed; $\Delta$ denotes the absolute
Pass@1 improvement over L1.}
\label{tab:memory}

\renewcommand{\arraystretch}{1.05}

\begin{tabular*}{0.76\columnwidth}
{@{\extracolsep{\fill}}lrrr@{}}
\toprule
\textbf{Abstraction Level} &
\textbf{Resolved} &
\textbf{Pass@1 (\%)} &
\textbf{$\Delta$ (pp)} \\
\midrule

L1
& 52/200
& 26.0
& -- \\

L1+L2
& 56/200
& 28.0
& +2.0 \\

L1+L2+L3
& \textbf{61/200}
& \textbf{30.5}
& \textbf{+4.5} \\

\bottomrule
\end{tabular*}
\end{table}

\section{Limitations}
\label{sec:limitations}

\tool has three main limitations. First, intermediate development history may be incomplete or unavailable, restricting reconstruction to the evidence preserved in the endpoint repair. Second, some repair histories span multiple problems and substantial execution and change context, making LLM-based
attribution and filtering more challenging. Third, validating each repair through the complete CI workflow would provide stronger feedback before proceeding to subsequent problems, but repeatedly executing these workflows is
computationally and time intensive. Moreover, CI workflows often depend on custom actions, services, configurations, and environments that are difficult to reproduce locally. \tool therefore relies on lightweight local validation when feasible rather than full CI execution after each repair step.

\section{Conclusion}
\label{sec:conclusion}
In this work, we introduced \tool, a framework for reconstructing reusable problem-level repair experience from entangled CI repair histories. \tool performs bidirectional change attribution by reasoning backward from endpoint patches and forward through development histories, using organized CI evidence to associate developer changes with individual CI problems. It represents each
resulting problem-repair experience at three abstraction levels that guide the repair of new CI failures. We evaluated {\tool} on the CI-Repair-Bench, where it consistently outperformed the studied baselines across multiple LLM backbones.

\bibliography{iclr2027_conference}
\bibliographystyle{iclr2027_conference}

\appendix
\appendix

\section{CI-Repair-Bench Characteristics}
\label{app:benchmark}

We provide additional characteristics of \textsc{CI-Repair-Bench} to quantify the repair and CI validation structure underlying our evaluation. We summarize the benchmark scale, the scope of validation and repair, the distribution of failure types, and the recurrence of the related repair experience.

\subsection{Benchmark Overview}
\label{app:benchmark-overview}

Table~\ref{tab:benchmark-overview} summarizes the scale, repair scope, CI validation characteristics, and failure-type distribution of \textsc{CI-Repair-Bench}. The benchmark contains 565 pull-request-level repair instances from 101 repositories. Each instance represents a repair trajectory from a failing revision to a developer-repaired revision. Because a repair may involve multiple CI problems, failure types are not mutually exclusive.

\subsection{Repair Recurrence}
\label{app:repair-recurrence}

We measure repair recurrence to quantify the extent to which
\textsc{CI-Repair-Bench} contains related repair experience. We consider two complementary views. \emph{Structural similarity} measures overlap in
normalized CI and repair attributes, including failure types, validation commands, dependencies, and modified files, using mean Jaccard similarity.
\emph{Lexical similarity} represents failure, workflow, and repair evidence using TF--IDF unigram and bigram features and measures their cosine similarity.

For each instance $i$, we define recurrence as its similarity to the most
similar eligible repair,
\[
R(i)=\max_{j\in\mathcal{C}_i} S(i,j),
\]
where $\mathcal{C}_i$ is the candidate repair set and $S$ denotes either
structural or lexical similarity. We report the mean nearest-neighbor score
under two scopes: \emph{within repository}, where candidates are restricted
to other repairs from the same repository, and \emph{overall benchmark},
where candidates may come from any repository.





The results show that related repair evidence recurs both within individual
repositories and across the benchmark, providing empirical motivation for
studying the reuse of prior repair experience.

\subsection{Historical and Evaluation Split}
\label{app:benchmark-split}

We partition the 565 benchmark instances chronologically within each
repository. For repositories with multiple instances, we assign the earliest $\lfloor 30\% \rfloor$ to the historical set and the remaining instances to evaluation, while retaining at least one instance in each set. Repositories with a single instance are assigned to evaluation. This procedure yields 157 historical repairs and 408 evaluation instances. The split is performed
before experience construction, ensuring that evaluation instances remain unseen and that historical repairs precede evaluation targets within each repository.

\begin{table*}[t]
\centering
\caption{Overview of \textsc{CI-Repair-Bench}. Total denotes the aggregate
quantity across all 565 repair instances. Mean, median, and maximum are
computed at the instance level. Failure types are not mutually exclusive;
the percentage denotes the fraction of instances containing each type.}
\label{tab:benchmark-overview}

\scriptsize
\renewcommand{\arraystretch}{1.08}

\begin{minipage}[t]{0.62\textwidth}
\centering
\textbf{Benchmark Characteristics}

\vspace{2pt}

\setlength{\tabcolsep}{2.2pt}
\begin{tabular*}{\linewidth}{
@{\extracolsep{\fill}}
llrrrr
@{}
}
\toprule
\textbf{Dimension} &
\textbf{Characteristic} &
\textbf{Total} &
\textbf{Mean} &
\textbf{Median} &
\textbf{Max} \\
\midrule

\multirow{3}{*}{Scale}
& Repair instances   & 565 & -- & -- & -- \\
& Repositories       & 101 & -- & -- & -- \\
& Failure categories &  12 & -- & -- & -- \\

\midrule

\multirow{4}{*}{CI validation}
& CI jobs      &  2,818 &  5.0 &  2 &  51 \\
& CI steps     & 15,561 & 27.5 & 14 & 557 \\
& Failed jobs  &  1,574 &  2.8 &  1 &  46 \\
& Failed steps &  1,636 &  2.9 &  1 &  46 \\

\midrule

\multirow{5}{*}{Repair scope}
& Commits        &   2,372 &   4.2 &  3 &    141 \\
& Files modified &   6,017 &  10.6 &  3 &    497 \\
& Lines changed  & 137,698 & 243.7 & 24 & 14,587 \\
& Lines added    &  80,725 & 142.9 & 15 &  7,503 \\
& Lines deleted  &  56,973 & 100.8 &  8 &  7,259 \\

\midrule

\multirow{2}{*}{Problem scope}
& Failure-type occurrences
& 1,907 & 3.38 & 3 & -- \\

& Multi-failure instances
& 473 & \multicolumn{3}{c}{83.7\% of instances} \\

\bottomrule
\end{tabular*}
\end{minipage}
\hfill
%
\begin{minipage}[t]{0.35\textwidth}
\centering
\textbf{Failure-Type Distribution}

\vspace{2pt}

\setlength{\tabcolsep}{2.2pt}
\begin{tabular*}{\linewidth}{
@{\extracolsep{\fill}}
lrr
@{}
}
\toprule
\textbf{Failure Type} &
\textbf{Instances} &
\textbf{\%} \\
\midrule

Linting               & 304 & 53.8 \\
Code Formatting       & 294 & 52.0 \\
Test Failure          & 279 & 49.4 \\
Type Checking         & 243 & 43.0 \\
Dependency Issues     & 243 & 43.0 \\
Runtime Error         & 180 & 31.9 \\
Configuration Error   & 172 & 30.4 \\
Doc/Docstring         &  81 & 14.3 \\
Assertion Error       &  48 &  8.5 \\
Package Install Error &  37 &  6.5 \\
Syntax Error          &  14 &  2.5 \\
Environment Error     &  12 &  2.1 \\

\bottomrule
\end{tabular*}
\end{minipage}

\end{table*}

\begin{table}[t]
\centering
\caption{Mean nearest-neighbor repair recurrence in
\textsc{CI-Repair-Bench}. Structural and lexical similarity are measured
independently.}
\label{tab:recurrence-results}

\footnotesize
\setlength{\tabcolsep}{8pt}
\renewcommand{\arraystretch}{1.02}

\begin{tabular}{@{}lrrr@{}}
\toprule
\textbf{Scope} &
\textbf{Instances} &
\textbf{Structural} &
\textbf{Lexical} \\
\midrule
Within repository & 507 & 0.436 & 0.811 \\
Overall benchmark & 565 & 0.445 & 0.796 \\
\bottomrule
\end{tabular}

\end{table}

\section{Additional Methodology Details}
\label{app:method-details}
\subsection{Dynamic Experience Reuse}
\label{app:dynamic-reuse}

This section provides implementation details for the dynamic experience-reuse
procedure described in Section~\ref{sec:experience-reuse}. \tool first decomposes the observed target CI failure into atomic problems, retrieves historical
experience separately for each problem, selects applicable experience, and uses it to construct repair guidance. All retrieval is based only on information available from the failing target revision.

\subsubsection{Target Problem Decomposition}
\label{app:target-decomp}

For each evaluation instance $x$, \tool first organizes the original failing CI logs and workflow configuration into a structured observation $O_x=\mathrm{Organize}(E_x)$. An LLM decomposes this observation into
atomic problems
\[
P_x=\langle p_{x1},\ldots,p_{xN_x}\rangle
\]
before repair begins. Each problem records its description, inferred root cause, concrete failure signals, affected files, failure type, validation
command, and available repository and workflow context.

This decomposition uses only the repository state, CI failure evidence(CI log), and workflow context available at the initial failing execution. The target developer patch, passing revision, and subsequent CI outcomes are not used. The corresponding baseline receives the same initial decomposition and problem-specific evidence.

\paragraph{Level-Specific Retrival Query Construction.}
\subsubsection{Level-Specific Query Construction}
\label{app:query}

For each target problem $p_{xj}$, \tool constructs a separate textual query
for each enabled abstraction level. The query fields reflect the degree of
abstraction represented by that level:

\begin{description}[leftmargin=*,topsep=2pt,itemsep=2pt]

\item[L1.] The query concatenates the repository, workflow filename, problem
description, root cause, affected files, failure types, and concrete failure
signals. This preserves the repository- and workflow-specific context of the
observed problem.

\item[L2.] The query contains the repository, problem description, root cause,
affected files, failure types, and failure signals. Workflow identity is
omitted, allowing retrieval at a broader repository-level scope.

\item[L3.] The query uses the abstract problem and root cause, failure types,
and generalized failure signals while excluding repository, workflow, and
file identifiers. This representation supports retrieval across repositories.

\end{description}

Thus, each observed problem produces three retrieval queries with
progressively less repository-specific context.

\subsubsection{Embedding Retrieval}
\label{app:retrieval}

\tool encodes queries and historical repair experience records at three abstraction levels using
\texttt{all-MiniLM-L6-v2}. Historical records are represented differently
at each level: L1 uses the repository, workflow, and problem descriptions
within the reconstructed repair episode; L2 uses the repository and
failure-identification summary; and L3 uses the failure pattern, abstract
problem description, and associated reasoning.

For level $\ell$, the target query and repair experience records at different abstractions are encoded with the
same encoder, and candidates are ranked by cosine similarity:
\[
s_{ij}^{(\ell)}
=
\frac{
\mathbf{q}_{xj}^{(\ell)\top}\mathbf{m}_{i}^{(\ell)}
}{
\|\mathbf{q}_{xj}^{(\ell)}\|_2
\|\mathbf{m}_{i}^{(\ell)}\|_2
}.
\]

Retrieval scope depends on the abstraction level. L1 is restricted to the same repository and workflow, L2 to the same repository, and L3 permits cross-repository retrieval. After applying these scope constraints, \tool retains the top $K=5$ records independently from each enabled level.

No minimum cosine-similarity threshold is used. The retrieved records are candidate experiences and are subsequently checked for applicability.

\paragraph{Applicability Selection.}
Semantic similarity alone does not determine whether a retrieved candidate
is applicable to the target problem. An LLM evaluates each candidate against
the current problem using its failure description, inferred root cause,
failure evidence, affected artifacts, validation context, and repair
applicability. Candidates judged relevant are retained for subsequent
enrichment. Repository-specific paths, commands, and assumptions from a
historical repair are treated as guidance and must be adapted to the target
repository. If no candidate is considered applicable, the current problem is
retained without historical guidance.

\subsubsection{Problem Enrichment}
\label{app:enrichment}

Selected historical experience is used to enrich the current problem before
repair. The enrichment combines the evidence observed in the current CI
failure with applicable information from the retrieved experience. Depending
on the retrieved level, this information may include prior root causes,
affected artifacts, concrete repair actions, repair strategies, applicability
conditions, validation guidance, and relevant problem relationships.
Information originating from historical repairs is retained as guidance
rather than treated as evidence that the same condition necessarily holds in
the target repository. When no applicable historical experience is available,
the original problem representation is retained unchanged.

\paragraph{Historical Dependency Information.}
During offline reconstruction, \tool infers dependency relations among
problems within each historical repair. A relation $p_i \rightarrow p_j$
indicates that the available repair evidence supports resolving $p_i$ before
$p_j$. The resulting dependency graph is topologically ordered so that
prerequisite repairs precede the problems that depend on them. These
relations preserve the structure and repair order of historical
multi-problem episodes and can provide evidence about problems that may not
yet be visible in the target CI execution.

\paragraph{Recurring Problem Recovery.}
In addition to the problems directly observed in the target CI failure,
\tool uses the retrieved L1 and L2 experience to identify problem patterns
that recur in historical repairs from the same repository and workflow.
For each target problem, the applicable L1 and L2 records identify relevant
historical repair instances. \tool examines the reconstructed problems across
these instances and groups similar problems to estimate their recurrence
within the corresponding repository--workflow scope. A problem pattern is
considered recurring when it appears in at least 50\% of the relevant
historical instances. Recurring patterns that are not already represented in
the target problem set are retained as candidate problems for the current
repair. Because these problems are inferred from historical recurrence rather
than directly observed in the current CI execution, they are treated as
historically supported repair hypotheses and are subsequently consolidated
with the observed problems.

\paragraph{Problem Clustering and Deduplication.}
To identify recurring problems across historical repairs, \tool groups
similar reconstructed problems using a weighted similarity score over the
problem description, root cause, affected files, failure type, repair
strategy, and file-type compatibility:
\[
S(p_i,p_j)
=
0.30S_{\mathrm{problem}}
+0.20S_{\mathrm{root}}
+0.20S_{\mathrm{file}}
+0.15S_{\mathrm{type}}
+0.10S_{\mathrm{fix}}
+0.05S_{\mathrm{filetype}}.
\]
Problem-description similarity receives the largest weight, while root-cause
and affected-file similarity provide additional evidence that two records
represent the same or a closely related problem. Failure type, repair
strategy, and file-type compatibility provide supporting signals. Problems
with $S(p_i,p_j)\geq0.50$ are assigned to the same cluster. A cluster is
considered recurring when it appears in at least 50\% of the historical
issues within the corresponding repository--workflow scope. The clustering
threshold is specific to historical problem grouping and is independent of
the cosine-similarity retrieval described in
Appendix~\ref{app:retrieval}.

\subsubsection{Sequential Repair Protocol}
\label{app:seq-repair}

The problem set to be repaired is determined before patch generation from the
observed CI problems together with any additional problems recovered from
applicable historical experience. \tool does not use benchmark-provided
future CI outcomes to extend this problem set during repair.

The resulting problems are processed sequentially in the same working tree.
Each problem is handled through a fresh agent interaction, while modifications
made for earlier problems remain in the repository and are therefore visible
to subsequent interactions. Where feasible, the agent may execute the
problem-specific validation command and use its output to refine the current
repair. Such execution feedback is used within the current repair interaction
and does not automatically introduce a new problem into the planned repair
sequence.

After all planned problems have been processed, the cumulative patch is
evaluated independently by re-executing the complete CI workflow. An instance
is counted as resolved only when the complete workflow passes.

The corresponding baseline receives the same initial problem decomposition
and problem-specific CI evidence but no retrieved historical repair
experience.\section{Prompt Templates}
\label{app:prompts}

\subsection{Historical Repair Reconstruction Prompts}
\label{app:historical-reconstruction-prompts}

\subsubsection{Endpoint Repair Reconstruction}
\label{app:endpoint-reconstruction}

Endpoint reconstruction is performed offline on historical successful repairs
used to construct repair experience. Given the observed CI failure, the completed developer patch, and the CI workflow context, \tool reconstructs the atomic problems addressed by the repair through three stages:
validation-step classification, atomic-problem extraction, and problem
consolidation. This process is applied only to historical repair instances;
ground-truth patches from evaluation instances are never used. The following
templates show the prompts used at each stage, with placeholders in braces
replaced by instance-specific evidence at runtime.
\paragraph{Stage 1: Validation-Step Classification.}
The first stage assigns each historical patch change to the CI validation step that would detect or require the corresponding repair.

\begin{lstlisting}[
style=promptbox,
caption={Prompt template for endpoint-view validation-step classification.},
label={lst:backward-classification}
]
TASK
Assign each change in the historical repair to the CI validation
step that would detect or require it.

INPUT

CI failure context:
{ci_failure_context}

Files visible in the CI failure logs:
{ci_visible_files}

Ordered CI validation sequence:
{validation_sequence}

Changed files and before/after changes:
{structured_patch_chunk}

Optional file-dependency evidence:
{file_dependency_context}

INSTRUCTIONS

Analyze the CI evidence, validation sequence, and historical
before/after changes jointly.

For each changed file:

1. Determine which CI validation step would detect or require
   the change.

2. Record the corresponding validation order and command.

3. Identify the failure type, specific issue type, and nature
   of the change.

4. Preserve the concrete before/after repair information,
   including relevant packages, configuration values, APIs,
   symbols, or source changes.

5. Mark the change as observed if its file is visible in the
   failing CI evidence; otherwise mark it as hidden.

6. When file-dependency evidence is available, determine whether
   related changes represent a supported causal adaptation.

Base the classification on the actual repair change and CI
context, not only on file type or extension. Do not infer failures
or dependencies that are unsupported by the supplied evidence.

OUTPUT (JSON)

[
  {
    "validation_order": <integer>,
    "validation_cmd": "<validation command>",
    "failure_type": "<failure category>",
    "issue_type": "<specific issue>",
    "change_type": "<code|dependency|config>",
    "change_scope_summary": [
      "<subject: before -> after>"
    ],
    "visibility": "<primary|hidden>",
    "files": ["<path>", "..."],
    "is_cascading": <boolean>,
    "dependency_type": "<relationship or empty string>",
    "cascade_explanation": "<explanation or empty string>"
  }
]

Every changed file must be accounted for exactly once. Preserve
the concrete repair evidence and return only evidence-supported
classifications.
\end{lstlisting}

\paragraph{Stage 2: Atomic-Problem Extraction.}
Changes assigned to the same validation step and failure category are grouped deterministically. Each resulting validation group is then decomposed into
atomic repair problems.

\begin{lstlisting}[
style=promptbox,
caption={Prompt template for endpoint-view atomic-problem extraction.},
label={lst:backward-atomic}
]
TASK
Decompose the changes in the current validation group into
atomic CI repair problems.

INPUT

CI failure context:
{ci_failure_context}

Files visible in the CI failure logs:
{ci_visible_files}

Validation context:
- validation_order: {validation_order}
- validation_cmd: {validation_command}
- failure_type: {failure_type}
- issue_type: {issue_type}
- change_type: {change_type}
- dependency context: {dependency_context}

Before/after repair changes:
{changes}

Validation-group boundaries:
{validation_group_summary}

INSTRUCTIONS

Infer the concrete CI problems repaired by the supplied
before/after changes.

For each problem:

1. Identify the failure signal or violated contract, the concrete
   problem, and its underlying root cause.

2. Identify the corresponding before-to-after repair and explain
   why the change resolves the problem.

3. Merge changes only when they represent the same failure,
   root cause, and compatible repair strategy. Keep changes
   separate when these differ, even if they affect the same file
   or validation step.

4. Keep problems within their assigned validation boundary.
   Changes associated with different validation steps must remain
   separate.

5. For dependency and configuration changes, preserve the
   concrete package, key, value, constraint, or operation needed
   to characterize the repair. Do not infer unsupported
   compatibility or dependency claims.

6. Mark a problem as "primary" when at least one affected file is
   visible in the failing CI evidence; otherwise mark it as
   "hidden".

7. Preserve supported causal information when the supplied
   dependency evidence indicates that the change is cascading.

OUTPUT (JSON)

{
  "atomic_problems": [
    {
      "problem_id": <integer>,
      "validation_order": <integer>,
      "validation_cmd": "<validation command>",
      "failure_type": "<failure category>",
      "issue_type": "<specific issue>",
      "problem": "<concrete problem and scope>",
      "root_cause": "<underlying cause>",
      "how_fixed": "<before-to-after repair>",
      "why_fix_works": "<repair rationale>",
      "affected_files": ["<path>", "..."],
      "problem_type": "<primary|hidden>",
      "is_cascading": <boolean>,
      "dependency_type": "<relationship or empty string>",
      "cascade_explanation": "<explanation or empty string>"
    }
  ]
}

Return an empty "atomic_problems" list when the supplied evidence
does not support a valid repair problem. Return JSON only.
\end{lstlisting}

\paragraph{Stage 3: Problem Consolidation.}
Semantically similar atomic problems are grouped as candidates, after which
the following prompt determines whether they represent the same repair problem
or should remain separate.

\begin{lstlisting}[
style=promptbox,
caption={Prompt template for endpoint-view problem consolidation.},
label={lst:backward-consolidation}
]
TASK
Given a cluster of candidate problems from the same validation,
group only problems that represent the same repair problem.

INPUT

Candidate problems:
{candidate_problems}

INSTRUCTIONS

Merge only if all of the following hold:
- same kind of failure;
- same underlying root cause;
- same or equivalent repair strategy; and
- no causal dependency exists between the candidate problems.

Keep problems separate if their root causes, violated contracts,
repair strategies, package operations, configuration operations,
or complexity differ.

OUTPUT (JSON)
{
  "problems": [
    {
      "merged_problems": [<input problem IDs>],
      "problem": "<unified problem description>",
      "root_cause": "<unified root cause>",
      "how_fixed": "<unified repair operation>",
      "why_fix_works": "<why it resolves all included cases>"
    }
  ]
}

Every input problem ID must occur in exactly one output group.
Preserve the union of all exact package, configuration, API,
symbol, and before/after details.
\end{lstlisting}

\subsubsection{Development-History Reconstruction Prompts}
\label{app:development-reconstruction-prompts}

Development-history reconstruction is performed offline on historical
successful repairs used to construct repair experience. It analyzes the
development trajectory from the failing revision to the successful revision
to recover how CI-relevant problems and repairs evolve across intermediate
developer changes. The procedure consists of four stages: CI-relevant file
selection, commit-level event extraction, cross-commit repair consolidation,
and dependency and repair-order reconstruction. The following templates
provide the prompts used at each stage, with placeholders in braces replaced
by instance-specific evidence at runtime.

\paragraph{Stage 1: CI-Relevant File Selection.}
For each commit in the historical trajectory, the first stage identifies
changed files whose modifications directly or indirectly affect a CI
validation.

\begin{lstlisting}[
style=promptbox,
caption={Prompt template for development-view CI-relevant file selection.},
label={lst:forward-file-selection}
]
TASK
Select changed files that are relevant to the CI validation
process.

INPUT

Commit metadata:
{commit_metadata}

Changed files:
{changed_files}

Commit diff:
{commit_diff}

Structured CI failure:
{structured_ci_failure}

Ordered CI validation sequence:
{validation_sequence}

Repository relationship context:
{relationship_context}

INSTRUCTIONS

Select every changed file whose modifications can directly or
indirectly affect a CI setup, installation, validation,
configuration, dependency, source, test, packaging, build,
generated-artifact, or documentation step.

For each candidate:
1. Identify the CI command or validation that consumes or checks it.
2. Inspect the actual changed hunks rather than relying on the
   filename or extension alone.
3. Include indirect effects supported by repository relationships,
   such as imports, calls, tests, configuration, generation,
   packaging, or dependency use.
4. Consider later validation steps even when they are absent from
   the initial failure log, since an earlier failure may prevent
   their execution.

Group files only when they correspond to the same validation
command, failure type, issue type, root cause, and reason for
CI relevance.

OUTPUT (JSON)
{
  "selected_groups": [
    {
      "files": ["<repository-relative path>", "..."],
      "failure_type": "<validation family>",
      "issue_type": "<specific issue family>",
      "validation_cmd": "<exact command or empty string>",
      "reason": "<why these changes affect the validation>"
    }
  ]
}

Every selected file must occur in the supplied commit diff.
If no changed file is relevant, return:
{"selected_groups": []}
\end{lstlisting}

\paragraph{Stage 2: Commit-Level CI Event Extraction.}
The selected changes are analyzed in trajectory order to identify the
CI-relevant events associated with each commit. The prompt distinguishes
problems that are introduced, repaired, replaced, or supported by the
commit while grounding each event in the observed repository changes.

\begin{lstlisting}[
style=promptbox,
caption={Prompt template for development-view commit-level CI event extraction.},
label={lst:forward-event-extraction}
]
TASK
Extract the CI-relevant events represented by the current
commit.

INPUT

Trajectory:
- failing commit: {sha_fail}
- successful commit: {sha_success}
- current commit: {commit_sha}
- position: {index}/{total_commits}

Commit metadata:
{commit_information}

CI-relevant changes:
{ci_relevant_diff}

Relevant file groups:
{selected_groups}

Observed CI failure:
{structured_ci_failure}

CI execution metadata:
{ci_metadata}

Relevant validation steps:
{validation_sequence}

INSTRUCTIONS

Analyze the selected changes and identify distinct events that
affect the listed CI validations.

For each event, determine whether the commit:
- introduces a CI problem;
- repairs a CI problem;
- replaces one problem with another; or
- provides supporting changes for another repair.

Use the commit changes as the primary evidence. CI execution
metadata provides additional evidence when available.

Consider direct effects on validated artifacts and indirect
effects supported by repository relationships, including source,
test, configuration, dependency, packaging, and build
relationships.

Group changes into one problem only when they represent the same
CI problem, root cause, and repair behavior. Preserve concrete
package, configuration, API, and before/after information when
needed to characterize the event.

Do not infer failures, compatibility constraints, or repair
effects unsupported by the supplied evidence.

OUTPUT (JSON)

{
  "commit_sha": "<current commit>",
  "problems": [
    {
      "files": ["<path>", "..."],
      "failure_type": "<failure category>",
      "issue_type": "<specific issue>",
      "problem": "<CI problem represented by the event>",
      "root_cause": "<underlying cause>",
      "changes_made": "<change introduced by this commit>",
      "introduced": <true|false>,
      "fixed": <true|false>,
      "fix_strategy": "<repair strategy, if applicable>",
      "why_this_fix_works": "<repair rationale>",
      "current_failed_jobs": ["<observed failed job>"],
      "current_fixed_jobs": ["<observed fixed job>"],
      "validation_cmd": "<validation command>"
    }
  ]
}

Use empty arrays when job-level CI evidence is unavailable.
Return an empty "problems" list when the commit contains no
supported CI-relevant event.
\end{lstlisting}
\paragraph{Stage 3: Cross-Commit Repair Consolidation.}
Commit-level events are grouped by validation command and failure type and ordered according to the CI validation sequence. Events across commits are
then analyzed jointly to determine whether they represent incremental parts of the same repair or distinct problems associated with the same validation.
The resulting events are consolidated into atomic problem--repair records.

\begin{lstlisting}[
style=promptbox,
caption={Prompt template for development-view trajectory consolidation.},
label={lst:forward-consolidation}
]
TASK
Analyze the grouped commit events and reconstruct the atomic CI
repair problems represented by this part of the trajectory.

INPUT

Grouped commit events:
{grouped_commit_events}

CI validation sequence:
{validation_sequence}

Validation context:
- validation_order: {validation_order}
- validation_cmd: {validation_command}
- failure_type: {failure_type}

INSTRUCTIONS

Use the commit events and available repository relationships to
identify the atomic repair problems represented by the trajectory.

Merge events only when they have:
- the same validation command;
- the same failure signal or violated contract;
- the same root cause;
- the same repair family; and
- compatible before/after behavior.

Keep events separate when they have different root causes,
validation contracts, repair strategies, or introduced/fixed
behavior, or when one problem causally depends on another.

PRIMARY/HIDDEN CLASSIFICATION

Set problem_type to "primary" when the problem appears in the
initial CI failure evidence.

Set problem_type to "hidden" when the problem is supported by the
historical trajectory and validation sequence but is absent from
the initial failure evidence.

OUTPUT (JSON)
{
  "atomic_problems": [
    {
      "problem_id": 1,
      "validation_order": <integer>,
      "validation_cmd": "<exact command>",
      "failure_type": "<failure category>",
      "issue_type": "<specific semantic issue>",
      "problem": "<what failed and its scope>",
      "root_cause": "<why the failure occurred>",
      "how_fixed": "<repair across the commit trajectory>",
      "why_fix_works": "<why the resulting state satisfies CI>",
      "affected_files": ["<path>", "..."],
      "problem_type": "<primary|hidden>",
      "is_cascading": <true|false>,
      "dependency_type": "<relationship or empty string>",
      "cascade_explanation":
        "<evidence-supported explanation or empty string>"
    }
  ]
}

Every affected file must come from the grouped commit events.
Do not invent files, validators, commands, or causal relations.
If the evidence does not support an atomic problem, return:
{"atomic_problems": []}
\end{lstlisting}

\subsubsection{Cross-View Reconciliation Prompts}
\label{app:cross-view-reconciliation-prompts}

Cross-view reconciliation integrates the atomic problem--repair records
independently produced by endpoint reconstruction and development-history
reconstruction. It does not repeat either reconstruction process. Instead,
the two sets of records are normalized and compared to identify overlapping,
complementary, conflicting, and distinct repair problems. Endpoint-derived
records are assigned identifiers $\mathrm{E}1,\mathrm{E}2,\ldots$, and
development-derived records are assigned identifiers
$\mathrm{D}1,\mathrm{D}2,\ldots$.

Before LLM-based reconciliation, candidate clusters are constructed from the
normalized records using their problem descriptions, root causes, repair
operations, failure and issue types, validation commands, and affected files.
The resulting clusters are processed in three stages:
(1) cross-view problem adjudication, (2) complementary evidence synthesis
when required, and (3) global problem selection and deduplication. The
following templates provide the prompts used at each stage, with placeholders
in braces replaced by instance-specific evidence at runtime.

\paragraph{Stage 1: Cross-View Problem Adjudication.}
For each candidate cluster, \tool compares the problem--repair records
reconstructed from the development and endpoint views and determines whether they
represent the same problem, provide complementary evidence, conflict, or
correspond to distinct repair problems.

\begin{lstlisting}[
style=promptbox,
caption={Prompt template for cross-view problem adjudication.},
label={lst:bidirectional-adjudication}
]
TASK
Analyze the forward and backward problem--repair records in the
candidate cluster.

INPUT

Candidate cluster:
{candidate_problems}

CI context:
- failed jobs: {failed_jobs}
- failure reasons: {failure_reasons}
- error types: {error_types}
- workflow: {workflow_name}

INSTRUCTIONS

Partition every source problem into exactly one decision group.
For each group, choose one action:

SELECT
The records describe the same problem and repair, or one is a
more complete version of the other. Select the best-supported
source record without rewriting it.

MERGE
The records describe the same underlying problem but contain
complementary, non-overlapping evidence. Merge them only when
combining the evidence produces a more complete problem without
introducing unsupported information.

CONFLICT
The records refer to the same problem but contain incompatible
repair operations or contradictory explanations.

SEPARATE
The records describe distinct repair problems and should remain
independent.

DECISION RULES

Use SELECT when:
- the repair operations are identical or near-identical;
- the records differ mainly in wording; or
- one record provides a more complete description of the same
  repair.

Use MERGE when:
- the records describe the same atomic problem;
- their evidence is complementary;
- one contributes failure, root-cause, or repair information
  absent from the other; and
- their repair operations do not conflict.

Use CONFLICT when:
- the records describe the same failure but prescribe
  incompatible repairs; or
- their root-cause explanations cannot simultaneously hold.

Use SEPARATE when:
- the failures, violated contracts, root causes, or repairs differ;
- one represents a distinct repair step; or
- similarity arises only from a shared file, validator, or
  terminology.

Identical or near-identical repairs must use SELECT rather than
MERGE. MERGE is reserved for genuinely complementary evidence.

OUTPUT (JSON)
{
  "analysis": "<brief cluster-level assessment>",
  "groups": [
    {
      "action": "<select|merge|conflict|separate>",
      "source_ids": ["F1", "B1"],
      "selected":
        "<source ID; required only for select>",
      "conflict_type":
        "<conflict category; required only for conflict>",
      "reason": "<evidence-supported decision>",
      "confidence": "<high|medium|low>"
    }
  ]
}

Every source_id in the candidate cluster must appear exactly once.
Do not generate or rewrite problem content in this stage.
\end{lstlisting}

For \texttt{SELECT}, the selected source record is retained directly.
For \texttt{SEPARATE}, the source records remain independent. Only groups
assigned \texttt{MERGE} require an additional synthesis step. If a source
record is omitted from the adjudication output, it is conservatively retained
as a separate problem.

\paragraph{Stage 2: Complementary Evidence Synthesis.}
When development-view and endpoint-view records describe the same atomic problem but provide
complementary evidence, \tool synthesizes them into a single problem record.
The synthesis preserves the repair operations and concrete evidence from both
views rather than generating a new repair hypothesis.

\begin{lstlisting}[
style=promptbox,
caption={Prompt template for complementary evidence synthesis.},
label={lst:bidirectional-synthesis}
]
TASK
Merge complementary descriptions of the same atomic CI repair
problem.

INPUT

Source problems:
{source_problems}

Reason for merge:
{merge_reason}

CI context:
{ci_context}

INSTRUCTIONS

Create one atomic problem that preserves all supported and useful
information from the source records.

REQUIREMENTS

1. Preserve every distinct failure aspect represented by the
   sources.
2. Preserve every distinct repair operation.
3. Use the union of all affected-file lists.
4. Preserve exact package names, configuration keys, versions,
   commands, symbols, and before/after values.
5. When two sources express the same fact at different levels of
   detail, retain the more specific supported description.
6. Do not introduce a file, repair, root cause, command, or
   technical detail absent from the source records.
7. Do not combine causally distinct repair problems.

Before producing the result, verify that every repair operation
from the source records is represented in how_fixed.

OUTPUT (JSON)
{
  "problem": "<unified problem description>",
  "root_cause": "<unified evidence-supported root cause>",
  "how_fixed": "<repair operations from the source records>",
  "why_fix_works": "<why the combined repair satisfies CI>",
  "affected_files": ["<union of source paths>"],
  "validation_cmd": "<most accurate supplied command>",
  "failure_type": "<failure category>",
  "issue_type": "<specific issue type>",
  "validation_order": <integer>,
  "source_ids": ["F1", "B1"]
}
\end{lstlisting}

The synthesized record retains its source identifiers as provenance. If
synthesis fails or produces unsupported content, the original source records
are retained rather than constructing a new problem.

\paragraph{Stage 3: Global Problem Selection and Deduplication.}
Cross-view adjudication operates on individual candidate clusters, so
duplicate or unsupported records may remain across clusters. \tool therefore
reviews the reconciled problem set jointly to retain CI-relevant problems,
remove unsupported records, and consolidate remaining global duplicates.

\begin{lstlisting}[
style=promptbox,
caption={Prompt template for global problem selection and deduplication.},
label={lst:bidirectional-global-selection}
]
TASK
Review the complete reconciled problem set for one historical
CI repair.

INPUT

Candidate problems:
{reconciled_candidate_problems}

CI context:
- failed jobs: {failed_jobs}
- failure reasons: {failure_reasons}
- error types: {error_types}
- relevant files: {ci_visible_files}
- validation sequence: {validation_sequence}
- workflow path: {workflow_path}

File and repository evidence:
{repository_evidence}

INSTRUCTIONS

For every candidate problem:

1. Determine whether it is directly relevant:
   it explains an observed failed job, command, error, or file.

2. Determine whether it is indirectly relevant:
   it is a concrete prerequisite or supporting repair for another
   retained problem.

3. Verify that the problem, root cause, repair operation, affected
   files, and validation command form a coherent and
   evidence-supported record.

4. Detect global duplicates. Two records are duplicates only when
   they describe the same atomic problem and repair. Sharing a
   file or validation command is insufficient.

5. Classify directly observed problems as primary and masked,
   prerequisite, or follow-on problems as hidden.

PROVENANCE RULES

- A backward-only problem originates from the historical repair
  patch and is retained unless it is unrelated to CI.
- A forward-only problem must be supported by the CI context,
  validation sequence, or repository evidence.
- A problem supported by both forward and backward decomposition
  is a strong candidate for retention.
- Distinct problems must not be removed merely because they share
  a validation command or affected file.

Reject a problem only when it is unsupported, unrelated to CI,
duplicated by a stronger record, or too vague to identify a
technical repair problem.

OUTPUT (JSON)
{
  "kept_problem_ids": [1, 2],
  "irrelevant_problem_ids": [3],
  "duplicate_groups": [
    {
      "keep_id": 1,
      "drop_ids": [4],
      "reason": "<same atomic problem and repair>"
    }
  ],
  "primary_problem_ids": [1]
}

Every supplied problem ID must be retained, marked irrelevant, or
included in a duplicate group. Do not rewrite or renumber the
problems.
\end{lstlisting}

The retained records form the reconciled historical problem set. Each record preserves its provenance as development-view-only, endpoint-view-only, or supported by both reconstruction views.

\subsection{Dependency Reconstruction Prompt}
\label{app:dependency-reconstruction-prompt}

After reconciliation, \tool analyzes the retained problems to identify
evidence-supported causal dependencies between repair steps. The following
prompt is used to infer candidate dependency edges; validation and repair-order
construction are described separately in the methodology details.

\begin{lstlisting}[
style=promptbox,
caption={Prompt template for causal dependency reconstruction.},
label={lst:dependency-reconstruction}
]
TASK
Infer causal dependencies among the retained problems of a
historical CI repair.

INPUT

Retained problems:
{final_problems}

CI context:
{ci_context}

Changed-file relationships:
{changed_file_graph}

Problem-level file relationships:
{problem_pair_graph}

Repair-change evidence:
{structured_diff}

INSTRUCTIONS

Compare the retained problems and identify evidence-supported
causal dependencies between their repairs.

For two problems A and B, infer A -> B only when:

1. A produces a concrete change in code, API, data, dependency,
   configuration, tool, or environment state;

2. B consumes that changed state or requires an adaptation to it;
   and

3. B's repair depends on the change introduced by A.

Use the following counterfactual check:

"If A's change were reverted, would B's repair still be required
for B's stated problem?"

If yes, treat the repairs as independent. If no, retain the edge
only when the supplied repository and repair evidence supports
the relationship.

EDGE DIRECTION

producer or prerequisite -> consumer or dependent

Consider relationships such as:
- caller -> callee or callee -> caller adaptations;
- reader -> writer dependencies;
- configuration -> consuming code or tool;
- dependency change -> source/API adaptation;
- source change -> test adaptation;
- build or installation change -> enabled validation.

Do not infer a dependency solely from:
- validation or temporal order;
- shared files, directories, or validators;
- problem-description similarity; or
- the fact that both repairs are required for CI success.

File relationships provide structural evidence, not dependency
evidence by themselves. Preserve the original problems and infer
only relationships supported by the supplied evidence.

OUTPUT (JSON)

{
  "dependencies": [
    {
      "from": <prerequisite_problem_id>,
      "to": <dependent_problem_id>,
      "dependency_type": "<causal relationship>",
      "producer_change": "<state introduced by the prerequisite>",
      "consumer_assumption": "<state required by the dependent>",
      "evidence_files": ["<repository-relative path>"],
      "reason": "<evidence-supported explanation>"
    }
  ]
}

Use only the supplied problem identifiers and historical repair
evidence. If no causal dependency is supported, return an empty
dependencies array.
\end{lstlisting}

\subsection{Prompts for constructing repair experience at three level abstractions}
\label{app:memory-prompts}

After historical repair reconstruction, \tool organizes the resulting
experience at three levels. L1 retains the concrete reconstructed repair
problems and their relationships. The following prompts construct L2
contextual repair strategies and L3 transferable repair patterns from this
concrete experience.

\subsubsection{Contextual Repair Strategy Prompt}
\label{app:l2-memory-prompt}

The L2 prompt transforms the organized concrete repair records (L1) into
contextual repair strategies. It groups related problems and summarizes how
they can be diagnosed, repaired, and validated while preserving relevant
dependency and repair-order information.

\begin{lstlisting}[
style=promptbox,
caption={Prompt template for L2 contextual repair strategy construction.},
label={lst:l2-memory-prompt}
]
TASK
Construct contextual repair strategies (L2) from the concrete
historical repair problems (L1).

INPUT

L1 repair problems:
{problems}

Available repair tools:
{automation_tools}

Repository and workflow context:
{repository_context}

INSTRUCTIONS

1. Analyze each L1 problem using its failure type, problem
   description, root cause, repair strategy, affected files,
   validation command, and dependency information.

2. Group problems that share the same underlying cause or belong
   to the same dependency-aware repair sequence. Keep independent
   problems as separate strategies.

3. For each group, construct a contextual repair strategy that
   specifies:
   - the failure or problems addressed;
   - the repair objective and rationale;
   - conditions under which the strategy applies;
   - observable diagnostic signals;
   - ordered repair actions;
   - applicable repair tools or manual actions;
   - verification guidance; and
   - relevant pitfalls.

4. Preserve dependency order when one repair is a prerequisite
   for another.

5. Ground every strategy in the supplied L1 evidence. Do not
   introduce unsupported failures, repair actions, or dependencies.

OUTPUT

Return structured L2 memory containing:
- a summary of the identified failures; and
- contextual repair strategies with applicability conditions,
  diagnostic signals, ordered actions, verification guidance,
  and pitfalls.
\end{lstlisting}

\subsubsection{L3 Transferable Repair Pattern Prompt}
\label{app:l3-memory-prompt}

The following prompt constructs transferable repair pattern (L3) from the
concrete L1 repair evidence and contextual L2 strategies. It abstracts
repository-specific experience into reusable repair patterns while preserving
the failure mechanism, diagnostic signals, repair approach, verification
guidance, and relevant repair relationships.

\begin{lstlisting}[
style=promptbox,
caption={Prompt template for L3 transferable repair memory construction.},
label={lst:l3-memory-prompt}
]
TASK
Construct transferable repair patterns (L3) from the concrete
repair evidence (L1) and contextual repair strategies (L2).

INPUT

Historical repair context:
- repository: {repo}
- workflow: {workflow}
- changed files: {changed_files}

L1 repair problems:
{problems}

L2 repair strategies:
{repair_strategies}

INSTRUCTIONS

Extract 1--3 repair patterns that can transfer beyond the source
repository.

For each pattern:

1. Identify the underlying failure mechanism from the L1 root
   causes and repair evidence.

2. Generalize repository-specific problems and repairs into a
   reusable failure pattern and repair approach.

3. Remove or abstract repository-specific names, exact paths,
   commands, and local conventions while preserving useful
   artifact types, failure signals, validator types, and repair
   mechanisms.

4. Use the L2 strategies to derive:
   - applicability conditions;
   - observable diagnostic signals;
   - the generalized repair approach;
   - ordered repair actions; and
   - verification guidance.

5. Preserve relevant automation guidance when it forms part of
   the repair strategy.

6. Preserve concrete L1 examples as supporting evidence for the
   generalized pattern.

7. Use supported L1 dependency relationships to capture relevant
   prerequisite or follow-on repair relationships without making
   the pattern depend on the exact validation sequence of the
   source repository.

Do not introduce a generalized pattern, repair action, or
dependency that is unsupported by the supplied L1 and L2
experience.

OUTPUT

Return 1--3 transferable repair patterns. Each pattern should
contain:
- failure type and generalized failure pattern;
- underlying failure mechanism;
- applicability conditions;
- observable diagnostic signals;
- generalized repair approach;
- ordered repair steps;
- verification guidance;
- supporting historical evidence; and
- relevant dependent or follow-on changes.

The resulting pattern must remain applicable beyond the source
repository while remaining grounded in the historical repair
evidence.
\end{lstlisting}

\subsection{Problem-Specific Repair Context}
\label{app:repair-context}

For each target problem, \tool constructs a problem-specific repair context
from the observed CI evidence and the retrieved historical experience. This
context is supplied to the patch-generation agent together with the current
repository state. The core context is shared across mini-SWE-Agent and Codex;
agent-specific formatting and execution instructions are described below.

\begin{lstlisting}[
style=promptbox,
caption={Problem-specific repair context supplied by \tool to the repair agent.},
label={lst:repair-context}
]
Problem {problem_index}/{total_problems}

Failure Type:
{failure_type}

Issue Type:
{issue_type}

Problem:
{problem_description_and_error_signals}

Root Cause:
{root_cause}

Affected Files:
{affected_files}

Repair Plan:

Approach:
{repair_strategy_summary}

Action Steps:
{repair_actions}

Validation Command:
{validation_command}

Pitfalls / Applicability Constraints:
{applicability_constraints}
\end{lstlisting}

The historical repair information is treated as guidance rather than as a
direct patch: the repair agent is instructed to verify its applicability
against the current repository and CI evidence before making changes.

\subsection{Repair Agent Prompts}
\label{app:agent-prompts}

\subsubsection{mini-SWE-Agent Repair Prompt}
\label{app:miniswe-prompt}

mini-SWE-Agent receives the problem-specific context in
Listing~\ref{lst:repair-context} together with a fixed repair instruction.
Problems are processed sequentially in the same working tree, allowing edits
from preceding repair steps to accumulate.

\begin{lstlisting}[
style=promptbox,
caption={Repair instruction template for mini-SWE-Agent.},
label={lst:miniswe-repair-prompt}
]
SYSTEM
You are an expert software engineer specializing in CI failure
repair. Analyze the CI failure and produce a minimal patch that
resolves the problem.

INSTRUCTIONS

1. Analyze the supplied problem and inspect the relevant repository
   files before making changes.

2. Verify that the supplied repair guidance applies to the current
   repository and CI evidence.

3. Determine the appropriate repair target, including source, test,
   configuration, dependency, or other relevant artifacts.

4. Apply the smallest complete repair for the current problem.
   Preserve valid edits already made for preceding problems.

5. Run the supplied validation command, or the narrowest relevant
   validation, when feasible.

6. If validation exposes an error introduced by the repair,
   correct it and repeat the targeted validation.

7. Avoid unrelated modifications and do not modify CI workflow
   files.

OUTPUT

Leave the completed repair in the shared working tree. The
cumulative working-tree diff forms the final patch.
\end{lstlisting}

\subsubsection{Codex Repair Prompt}
\label{app:codex-prompt}

Codex receives the same problem-specific repair context in
Listing~\ref{lst:repair-context}, together with its agent-specific repair
instructions. Each problem is processed in the same repository checkout,
and Codex may iteratively inspect, edit, validate, and revise the repair
within an invocation.

\begin{lstlisting}[
style=promptbox,
caption={Repair instruction template for Codex.},
label={lst:codex-repair-prompt}
]
TASK
Repair the supplied CI problem in the current repository.

INSTRUCTIONS

1. Analyze the supplied CI evidence and inspect the current
   repository to determine the root cause.

2. Distinguish observed evidence from inferred repair hypotheses
   and confirm inferred problems before editing.

3. Verify that the supplied historical repair guidance applies
   to the current repository.

4. Implement the smallest complete repair while preserving valid
   changes from preceding problems.

5. Run the supplied validation command when available; otherwise,
   derive the narrowest relevant validation from the CI
   configuration.

6. Inspect validation feedback and revise the repair when the
   current change is incomplete or introduces a related error.

7. Do not weaken validation, modify unrelated files, or introduce
   unsupported dependency or configuration changes.

OUTPUT

Leave the completed repair in the working tree and report the
changed files and targeted validation performed.
\end{lstlisting}

\section{Experimental Setup Details}
\label{app:experimental-details}

This section provides the backbone-model specifications, repair-agent
configurations, and baseline settings used in our experiments.

\subsection{Backbone Models}
\label{app:backbone-models}

We evaluate \tool with MiniMax-M2.5 and DeepSeek-V4-Flash.
Table~\ref{tab:backbone-models} summarizes the model specifications relevant to our evaluation. Model characteristics and pricing follow the corresponding official documentation~\citep{minimax2026m25,deepseekai2026deepseekv4}.

\subsection{Backbone Models}
\label{app:backbone-models}

We evaluate \tool with MiniMax-M2.5 and DeepSeek-V4-Flash.
Table~\ref{tab:backbone-models} summarizes the model specifications and
inference settings used in our evaluation. Model characteristics and pricing
follow the corresponding model and API documentation.

For each model and agent comparison, the backbone and inference settings are
held fixed between \tool and the corresponding agent-only baseline. We use
temperature $0.0$ throughout the evaluation to reduce sampling variability
across experimental conditions.
For each model--agent comparison, the backbone and inference settings are
held fixed between \tool and the corresponding agent-only baseline.

\subsection{Patch Generation}
\label{app:patch-generation}

We use mini-SWE-Agent and Codex for patch generation. Both operate on the
failing repository revision and return the cumulative working-tree changes
after repair.

\subsubsection{mini-SWE-Agent}
\label{app:mini-swe-agent}

mini-SWE-Agent interacts with the repository through shell commands. Problems
within the same CI instance are processed sequentially in a shared working
tree, allowing edits from earlier repair steps to remain available to later
steps. For each problem, the agent inspects the failure evidence and current
repository state, applies a targeted repair, and runs the narrowest applicable
validation when feasible.
\begin{table}[t]
\centering
\caption{Backbone models and inference settings used in our evaluation.
Context limit denotes the maximum context configured in our experiments.
Prices are reported in USD per one million tokens.}
\label{tab:backbone-models}

\footnotesize
\setlength{\tabcolsep}{2.0pt}
\renewcommand{\arraystretch}{1.02}

\begin{tabular}{@{}lcccccccc@{}}
\toprule
\textbf{Model} &
\textbf{Release} &
\textbf{Params.} &
\textbf{\shortstack{Native\\Context}} &
\textbf{\shortstack{Context\\Limit}} &
\textbf{Temp.} &
\textbf{Input} &
\textbf{\shortstack{Cache\\Read}} &
\textbf{Output} \\
\midrule

MiniMax-M2.5
& Feb. 2026
& 229B
& 200K
& 200K
& 0.0
& \$0.15/M
& \$0.030/M
& \$1.20/M \\

DeepSeek-V4-Flash
& Apr. 2026
& \shortstack{284B /\\13B active}
& 1M
& 163,840
& 0.0
& \$0.03/M
& \$0.016/M
& \$0.32/M \\

\bottomrule
\end{tabular}

\end{table}

The repair-agent instruction used for mini-SWE-Agent is provided in
Appendix~\ref{app:miniswe-prompt}.

\subsubsection{Codex}
\label{app:codex-agent}

Codex provides a more iterative patch-generation procedure. Each problem is
handled by a separate invocation in the same working tree. Within an
invocation, Codex can inspect files, modify the repository, execute targeted
validation, and revise its repair based on the resulting feedback.

\begin{table*}[t]
\centering

\begin{minipage}[t]{0.48\textwidth}
\centering
\caption{Codex configuration used in our experiments.}
\label{tab:codex-config}

\footnotesize
\setlength{\tabcolsep}{4pt}
\renewcommand{\arraystretch}{1.02}

\begin{tabular}{@{}ll@{}}
\toprule
\textbf{Setting} & \textbf{Value} \\
\midrule
Backbone model    & MiniMax-M2.5 \\
Interface         & Codex CLI \\
Reasoning effort  & Medium \\
Sandbox           & Workspace-write \\
Wall-clock limit  & 3,600 s per problem \\
Repair execution  & Sequential, shared working tree \\
Validation        & Iterative within invocation \\
Final output      & Cumulative working-tree diff \\
\bottomrule
\end{tabular}
\end{minipage}
\hfill
%
\begin{minipage}[t]{0.48\textwidth}
\centering
\caption{Configuration of the ThinkRepair baseline.}
\label{tab:thinkrepair-config}

\footnotesize
\setlength{\tabcolsep}{4pt}
\renewcommand{\arraystretch}{1.02}

\begin{tabular}{@{}ll@{}}
\toprule
\textbf{Setting} & \textbf{Value} \\
\midrule
Knowledge source             & Historical repair instances \\
Retrieved demonstrations     & $k=2$ \\
Reasoning representation     & Chain-of-thought (CoT) \\
Planning model               & MiniMax-M2.5 \\
Planning temperature         & $0.7$ \\
Patch-generation agent       & mini-SWE-Agent \\
Patch-generation temperature & $0.0$ \\
\bottomrule
\end{tabular}
\end{minipage}

\end{table*}

The Codex repair instruction is provided in
Appendix~\ref{app:codex-prompt}. The same Codex configuration is used across
the agent-only and \tool runs; \tool additionally supplies the retrieved
problem-specific repair guidance.

\subsection{Baseline Configurations}
\label{app:baseline-configurations}

\subsubsection{Agent-Only Baseline}
\label{app:agent-only-baseline}

The agent-only baseline uses the same patch-generation agent, backbone, and
execution settings as the corresponding \tool configuration, but receives no
historical repair experience. It operates on the observed target failure,
current repository state, and workflow-derived validation information.

\subsubsection{ThinkRepair}
\label{app:thinkrepair}

We adapt ThinkRepair~\citep{yin2024thinkrepair} as an experience-based
baseline. Following ThinkRepair, we construct a knowledge pool from historical
repair instances and retrieve relevant knowledge as chain-of-thought (CoT)
demonstrations for the target problem. The model uses these demonstrations to
reason about the target failure and produce a step-by-step repair plan. We
then provide this plan to mini-SWE-Agent, which applies the repair to the
target repository and generates the final patch.




The ThinkRepair prompt follows the original CoT-based reasoning
framework~\citep{yin2024thinkrepair}; our adaptation uses the generated
reasoning as a repair plan for mini-SWE-Agent.

\section{Additional Experimental Analysis}
\label{app:additional-analysis}

\subsection{Overlap Across Reconstruction Strategies}
\label{app:reconstruction-overlap}

\begin{figure*}[t]
    \centering

    \begin{minipage}[t]{0.48\textwidth}
        \centering
        \includegraphics[width=\linewidth]{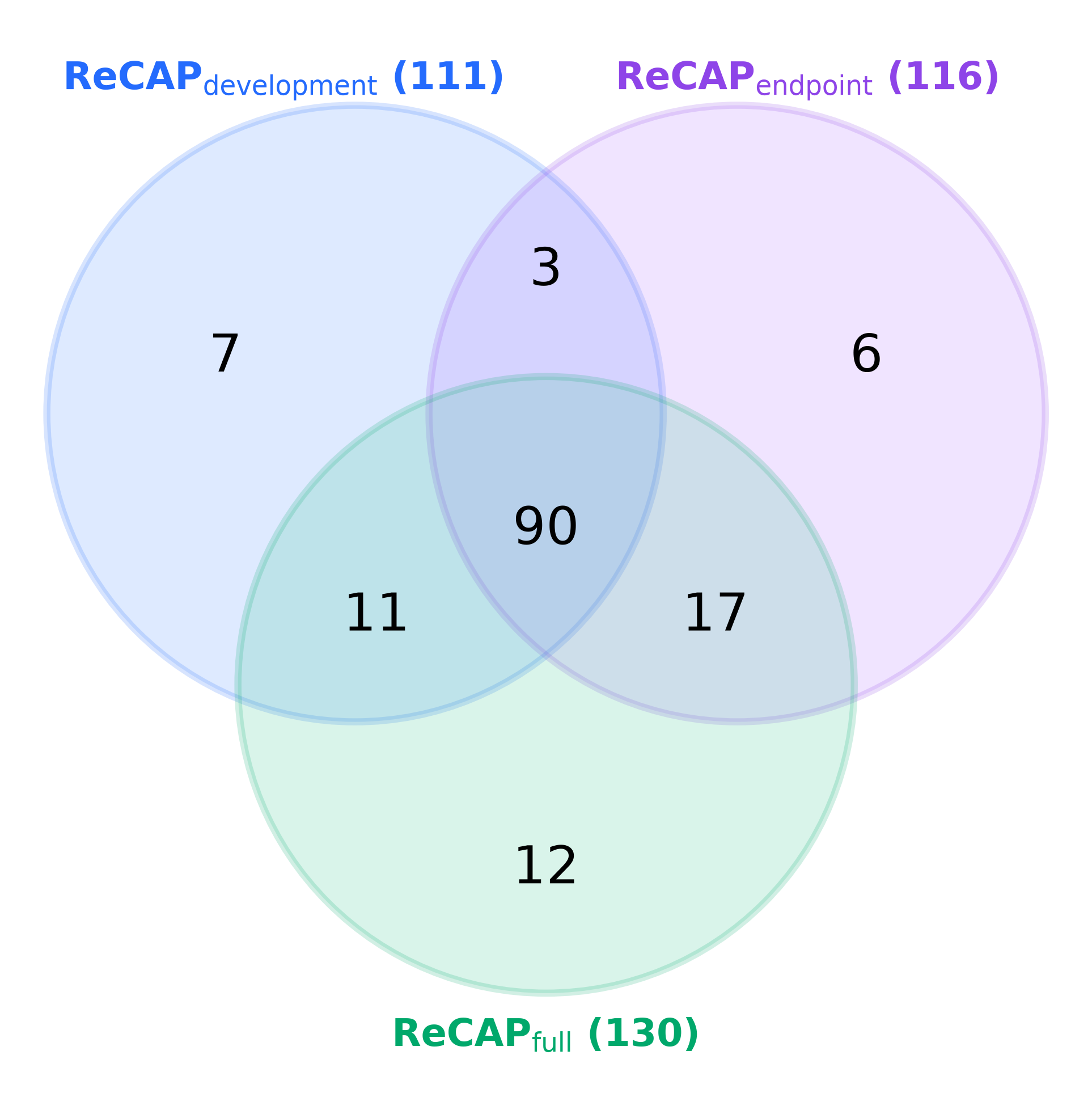}
        \par\smallskip
        \small (a) MiniMax-M2.5
    \end{minipage}
    \hfill
    \begin{minipage}[t]{0.48\textwidth}
        \centering
        \includegraphics[width=\linewidth]{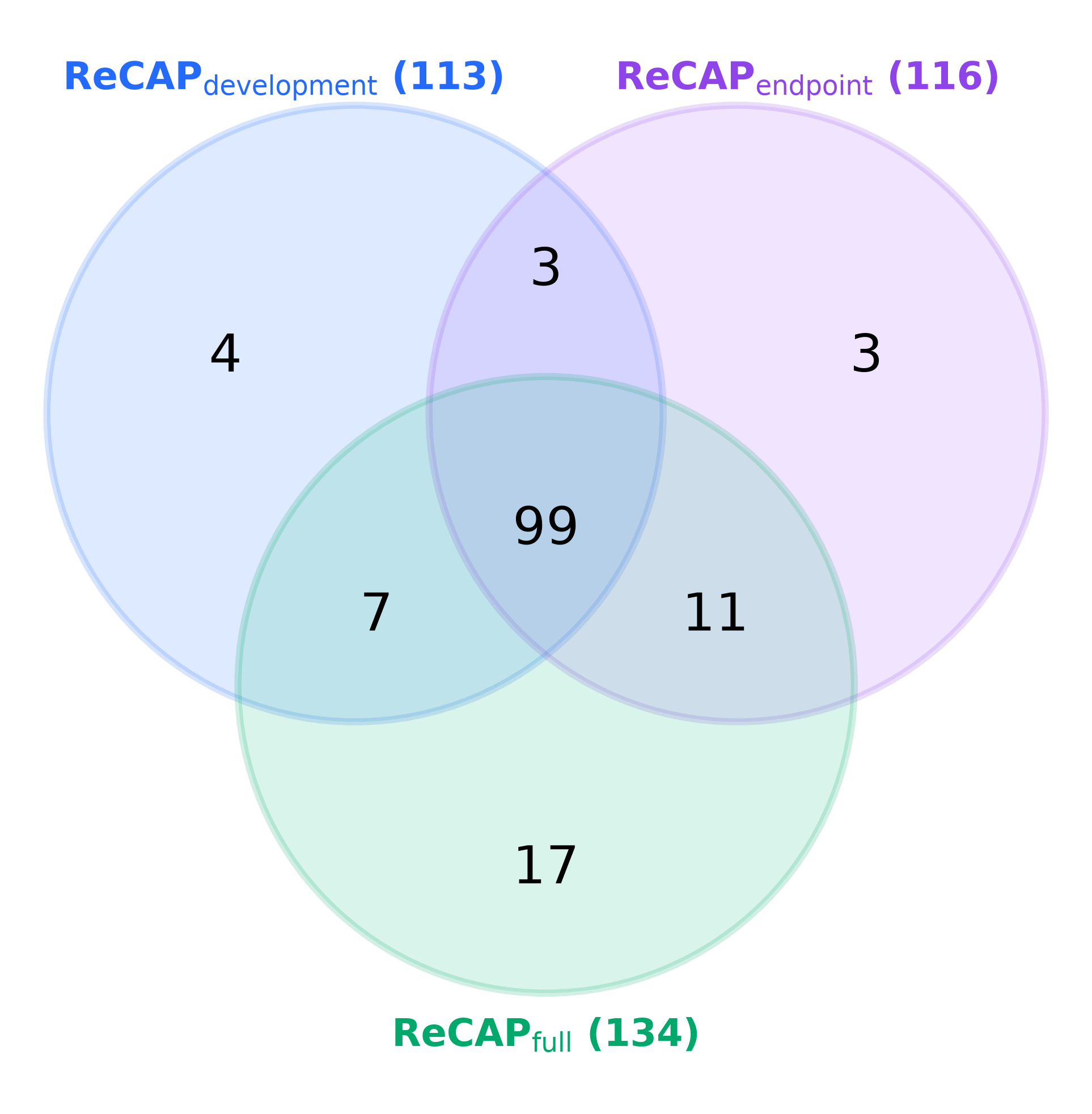}
        \par\smallskip
        \small (b) DeepSeek-V4-Flash
    \end{minipage}

    \caption{Overlap of resolved instances among the three \tool
    reconstruction configurations on the 408-instance evaluation set using
    mini-SWE-agent.}
    \label{fig:reconstruction-overlap}
\end{figure*}

Figure~\ref{fig:reconstruction-overlap} examines the overlap among instances
resolved by the endpoint, development, and full reconstruction strategies
with mini-SWE-agent. With MiniMax-M2.5, at least one strategy resolves 146
of the 408 evaluation instances (35.8\%), including 90 resolved by all three
configurations; the full configuration uniquely resolves 12 instances.
Similarly, with DeepSeek-V4-Flash, at least one strategy resolves 144
instances (35.3\%), including 99 resolved by all three configurations; the
full configuration uniquely resolves 17 instances. These unique successes,
together with the substantial overlap across configurations, show that the
endpoint and development reconstruction views provide complementary repair
information. This provides an instance-level view of the reconstruction
ablation in Table~\ref{tab:reconstruction}, complementing the aggregate
Pass@1 results.

\end{document}